\documentclass[%
superscriptaddress,
preprint,
 amsmath,amssymb,
 aps,
 prl
]{revtex4-2}

\usepackage{graphicx}
\usepackage{dcolumn}
\usepackage{bm}
\usepackage[version=4]{mhchem}
\usepackage{siunitx}
\usepackage{physics}
\usepackage{booktabs}  
\usepackage{afterpage}
\usepackage{placeins}
\usepackage{multirow}
\usepackage{dcolumn}
\newcolumntype{d}[1]{D{.}{.}{#1}} 
\newcolumntype{R}[1]{>{\raggedleft\arraybackslash$}p{#1}<{$}}
\DeclareSIUnit{\angstrom}{\ensuremath{\mathrm{\AA}}}

\begin{document}

\preprint{}

\title{Picocavity-Enhanced Raman Spectroscopy of Physisorbed H$_2$ and D$_2$ Molecules}

\author{Akitoshi Shiotari}
 \email{Corresponding author: shiotari@fhi-berlin.mpg.de}
\affiliation{
 Department of Physical Chemistry, Fritz-Haber Institute of the Max-Planck Society, Faradayweg 4-6, 14195 Berlin, Germany
}

\author{Shuyi Liu}
\altaffiliation[Current address: ]{Huazhong University of Science and Technology, Wuhan, China.}
\affiliation{
 Department of Physical Chemistry, Fritz-Haber Institute of the Max-Planck Society, Faradayweg 4-6, 14195 Berlin, Germany
}

\author{George Trenins}
\affiliation{
 Max-Planck-Institute for Structure and Dynamics of Matter, 22761 Hamburg, Germany
}

\author{Toshiki Sugimoto}
\affiliation{
 Institute for Molecular Science, National Institutes of Natural Sciences, 444-8585 Okazaki, Japan
}
\affiliation{
 The Graduate University for Advanced Studies, SOKENDAI, 240-0193 Hayama, Japan
}

\author{Martin Wolf}
\affiliation{
 Department of Physical Chemistry, Fritz-Haber Institute of the Max-Planck Society, Faradayweg 4-6, 14195 Berlin, Germany
}

\author{Mariana Rossi}
 \email{Corresponding author: mariana.rossi@mpsd.mpg.de}
\affiliation{
 Max-Planck-Institute for Structure and Dynamics of Matter, 22761 Hamburg, Germany
}

\author{Takashi Kumagai}
 \email{Corresponding author: kuma@ims.ac.jp}
\affiliation{
 Institute for Molecular Science, National Institutes of Natural Sciences, 444-8585 Okazaki, Japan
}
\affiliation{
 The Graduate University for Advanced Studies, SOKENDAI, 240-0193 Hayama, Japan
}

\date{\today}

\begin{abstract}
We report on tip-enhanced Raman spectroscopy of H$_2$ and D$_2$ molecules physisorbed within a plasmonic picocavity at 10 K.
The intense Raman peaks resulting from the rotational and vibrational transitions are observed at subnanometer gap distances of the junction formed by an Ag tip and an Ag(111) surface, where a picocavity-enhanced field plays a crucial role.
A significant redshift of the H-H stretch frequency is observed as the gap distance decreases, while the D-D stretch frequency is unaffected.
Density functional theory, path-integral molecular dynamics, and quantum anharmonic vibrational energy calculations suggest that this unexpected isotope effect is explained by a different molecular density between H$_2$ and D$_2$ on the surface.
\end{abstract}

\maketitle

The adsorption of hydrogen molecules on solid surfaces is the first step in 
permeation for fuel storage \cite{ockwig2007membranes}, catalytic hydrogenation \cite{zhang2019selective}, hydrogen embrittlement \cite{li2020review}, and nuclear-spin isomer conversion \cite{fukutani2013physisorption}.
To understand and control the elementary processes in such physical and chemical phenomena, hydrogen adsorption has been intensively studied in surface science \cite{christmann1988interaction,vidali1991potentials,bruch2007progress}.
Specifically, hydrogen, the lightest molecule, exhibits unique quantum effects that differ from those of other molecules when it is physisorbed on a surface through van der Waals interactions \cite{avouris1982observation,andersson1982observation,yu1983selective,sakurai1988ortho,anger1989adsorption,williams2002raman,fukutani2003photostimulated,niki2008mechanism,sugimoto2014effects,sugimoto2017inelastic}.  
However, only a few methods are available for characterizing such physisorption systems as the weak adsorption energy ($< 100$ meV) necessitates measurements at cryogenic temperatures.
Low-temperature scanning tunneling microscopy has been employed as a local characterization technique to investigate the dynamics of hydrogen molecules weakly adsorbed on surfaces \cite{gupta2005strongly,temirov2008novel,lotze2012driving,li2013rotational,natterer2013distinction,natterer2014resonant,therrien2015collective,liu2018enhanced,merino2018single,wang2022atomic,wang2023electrical}.
Several studies have observed characteristic features in conductance spectra, attributed to the rotational transition of molecular hydrogen in the scanning tunneling microscopy (STM) junction \cite{li2013rotational,natterer2013distinction,natterer2014resonant,therrien2015collective}.
However, the interpretation remains controversial;
the features within the corresponding energy range [30--60 meV ($\sim$240--480 cm$^{-1}$) for H$_2$] could be attributed to configuration switching \cite{gupta2005strongly,thijssen2006vibrationally,halbritter2008huge}, to a molecule--substrate stretching mode \cite{li2015symmetry}, or to multiple phonon excitation \cite{trouwborst2009bistable}. 
Additionally, the stretching mode has never been observed in these studies.

Raman spectroscopy is a promising technique for characterizing hydrogen molecules on surfaces, as it allows the observation of both rotational and vibrational transitions of homonuclear diatomic molecules.
Although the detection demands exceptional sensitivity due to the intrinsically small cross section, single-molecule detection can be achieved by tip-enhanced Raman spectroscopy (TERS) and surface-enhanced Raman spectroscopy (SERS) \cite{haran2010single,ding2016nanostructure,wang2020fundamental,itoh2023toward,hoppener2024tip}.
Recent studies have established that picocavities, atomic-scale protrusions present in plasmonic nanojunctions, play a crucial role in achieving extreme confinement of the electromagnetic field to angstrom scales \cite{benz2016single,urbieta2018atomic,hoppener2024tip}, resulting in a significant enhancement of Raman scattering \cite{liu2019resolving,liu2023inelastic,park2024atomic} and ultrahigh spatial resolution \cite{zhang2013chemical,lee2019visualizing,xu2021determining,litman2023jpcl}.
Yet, the exceptional sensitivity in TERS and SERS has been demonstrated primarily for organic molecules chemisorbed onto the substrate, where the enhancement benefits from chemical effects arising from the molecule--surface (and molecule--tip) interactions \cite{chen2023interpreting}.
Therefore, applying TERS and SERS to physisorbed systems has remained challenging.

\begin{figure*}
\includegraphics[width=16cm]{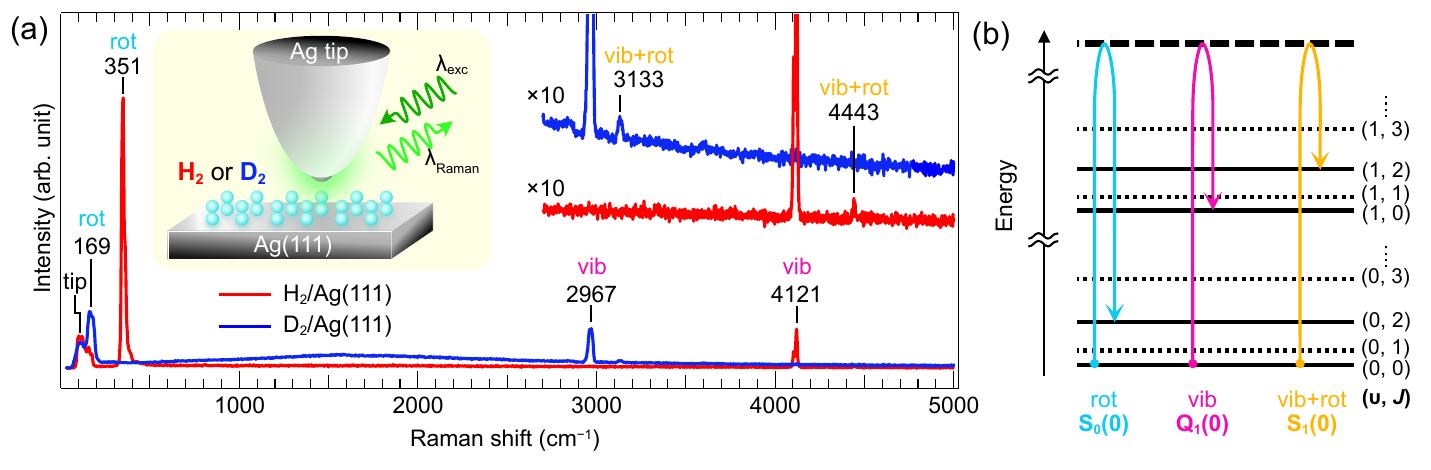}
\caption{\label{fig1}
(a) TERS spectra of H$_2$ (red) and D$_2$ (blue) on Ag(111) with a Ag tip at 10 K (sample bias voltage $V_\mathrm{s}$ = 10 mV, tunneling current $I_\mathrm{t}$ = 1.0 nA). 
The left inset shows a schematic of the experiment.
$\lambda_\mathrm{exc}$ and $\lambda_\mathrm{Raman}$ denote the wavelength of the incident laser and Raman scattering, respectively.
The right inset shows magnified spectra in the high wavenumber regions.
The peak originating from tip phonon modes is denoted as ``tip.''
(b) Energy diagram of the vibrational and rotational levels of an H$_2$ or D$_2$ molecule.
The solid lines represent the levels for \textit{para}-H$_2$ and \textit{ortho}-D$_2$, while the dotted lines are those for \textit{ortho}-H$_2$ and \textit{para}-D$_2$.
The dashed bold line represents a virtual state.
The arrows indicate Raman-active transitions referred to as ``rot,'' ``vib,'' and ``rot+vib.''
$\upsilon$ and $J$ denote the vibrational and rotational quantum numbers, respectively.
}
\end{figure*}

Here, we demonstrate that low-temperature TERS (LT-TERS) can detect hydrogen molecules physisorbed onto an Ag(111) surface with an Ag tip [Fig.~\ref{fig1}(a), inset], allowing for the identification of their rotational and vibrational states.
Furthermore, precise gap-distance control enables us to investigate not only the contribution of the picocavity field to the Raman scattering process but also the influence of tip proximity on the rotational and vibrational energies of the molecules through nontrivial nuclear quantum effects. 
Density functional theory (DFT), path-integral molecular dynamics, and quantum vibrational energy calculations suggest that the tip--molecule--surface potential energy profiles are critically affected by the intermolecular and tip--molecular interactions, leading to unusual isotope effects on vibrational frequencies.

The LT-TERS experiments were performed in an ultrahigh vacuum chamber [see Sec.~I-A of Supplemental Material (SM) \cite{SI}]. 
The 532-nm incident laser beam was linearly polarized along the tip axis and focused on the apex of a focused-ion-beam-sharpened Ag tip \cite{liu2019resolving,park2024atomic} using an \textit{in situ} parabolic mirror, and the scattering light was detected by an \textit{ex situ} spectrometer. 
We conducted DFT calculations \cite{blum+cpc2009,gavi+roadmap2023} of H$_2$/D$_2$ on a $4\times4$ Ag(111) (periodic) surface, employing the Perdew-Burke-Ernzerhof exchange-correlation functional and screened pairwise van der Waals interactions \cite{ruiz2012prl} (see SM Sec.~I-B \cite{SI}).
The preferred adsorption sites (fcc or hcp hollow) and binding energies ($\sim$30 meV) were in very good agreement with previous studies of isolated molecules on the surface \cite{kunisada2015hindered,smeets2021designing}.
Unless specified otherwise, we considered a coverage of 0.69 monolayer (ML), motivated by the results of path-integral molecular dynamics simulations at low temperature 
(see SM Sec.~I-B \cite{SI}).
At 0.69 ML coverage, the intermolecular attractive interaction increases the binding energy by 55 meV compared to the lowest coverage of 0.06 ML considered here (Fig.~S3).
Over the sample, we placed the Ag-tip model shown in SM Sec.~I-B \cite{SI} (see also Fig.~S4), which was also previously discussed in Refs.~\cite{litman2023jpcl,cirera2022charge,liu2023inelastic}.

H$_2$ or D$_2$ gas was introduced into the chamber where a clean Ag(111) surface was cooled at 10 K (see SM Sec.~I-A \cite{SI}). 
Regardless of laser illumination, no static hydrogen molecules on the surface were imaged with STM (Fig.~S1). 
In contrast, clear peaks were observed in TERS spectra under these conditions. 
Figure~\ref{fig1}(a) displays typical TERS spectra of H$_2$ and D$_2$ on Ag(111). 
For H$_2$, intense peaks were observed at 351 and 4121~cm$^{-1}$, along with a weak peak at 4443~cm$^{-1}$. 
Similarly, for D$_2$, intense peaks appeared at 169 and 2967~cm$^{-1}$, with a weak peak at 3133~cm$^{-1}$.
The three peaks can be assigned to rotational and vibrotational modes of hydrogen.
Note that broad or asymmetric peak shapes are artifacts in the wide wavenumber range spectra (see SM Sec.~I-A \cite{SI}).
The feature around 100 cm$^{-1}$ common to both spectra originates from phonon modes of the Ag tip \cite{liu2023inelastic,cui2023atomistic}, which is present even in the absence of hydrogen adsorption (Fig.~S2). 
Figure~\ref{fig1}(b) depicts the energy diagram of rotational and vibrational levels and allowed transitions under the Raman selection rule. 
The observed peaks, in order from lowest to highest, correspond to the lowest-energy rotational transition [$\mathrm{S}_0(0)$; ``rot'' labeled in Fig.~\ref{fig1}(a)], the lowest-energy vibrational transition [$\mathrm{Q}_1(0)$; ``vib''], and their combination [$\mathrm{S}_1(0)$; ``rot+vib''], respectively.
The three modes were also observed in high-resolution electron energy loss spectroscopy of hydrogen on a polycrystalline Ag film \cite{avouris1982observation} and a Cu(001) surface \cite{andersson1982observation} at approximately 10--12 K; however, better energy resolution ($<$10 cm$^{-1}$) in TERS than that of the electron energy loss spectroscopy ($>$28 cm$^{-1}$ \cite{andersson1982observation}) makes the redshifts from gas-phase hydrogen more discernible (Table~S4). 
We should also note that only the stable nuclear-spin isomer of each isotope, i.e., \textit{para}-H$_2$ and \textit{ortho}-D$_2$ [Fig.~\ref{fig1}(b)], is detected in TERS (Fig.~S2).
On the surface, the stable isomer is dominant due to the \textit{ortho}--\textit{para} conversion \cite{avouris1982observation,fukutani2013physisorption,kunisada2015hindered} (see SM Sec.~II-A \cite{SI}). 

The rotational transition at 351 cm$^{-1}$ for H$_2$/Ag(111) [Fig.~\ref{fig1}(a)] is slightly redshifted from 354 cm$^{-1}$ in the gas phase \cite{veirs1987raman}.
We calculate the H-H bond length to be 75.4 pm on Ag(111) (cf.~75.1 pm in the gas phase).
Considering a rigid-rotor approximation, the rotational transition energy of \textit{para}-H$_2$ is calculated to be 353 cm$^{-1}$ on the surface, compared to 356 cm$^{-1}$ in the gas phase, showing  good agreement with the experiment.
Notably, the H-H bonding distance of H$_2$ on Ag(111) varied by only 0.06 pm at the closest tip position with respect to simulations without any tip. 
This suggests that the hydrogen molecules on the surface and in the STM junction behave as nearly free rotors.

In the TERS spectrum [Fig.~\ref{fig1}(a)], the vibrational frequency of 4121~cm$^{-1}$ exhibits a redshift relative to the gas-phase value of 4161~cm$^{-1}$ \cite{veirs1987raman}.
The vibrational transition energy is calculated to be 4088 cm$^{-1}$ on the surface (without the Ag tip), after applying a scale factor to the calculations performed in the harmonic approximation (see SM Sec.~I-C \cite{SI}; see also Fig.~S8 for a simulated LT-TERS spectrum \cite{litman2023jpcl}).
The slight discrepancy between the experimental and calculated frequencies is consistent with the typical spurious softening of the Perdew-Burke-Ernzerhof functional. 
Compared to the calculated gas-phase value, the vibrational frequency shows a redshift of $\sim$73 cm$^{-1}$ upon the adsorption, which is larger than, but in reasonable agreement with the experiment.

\begin{figure}
\includegraphics[width=8cm]{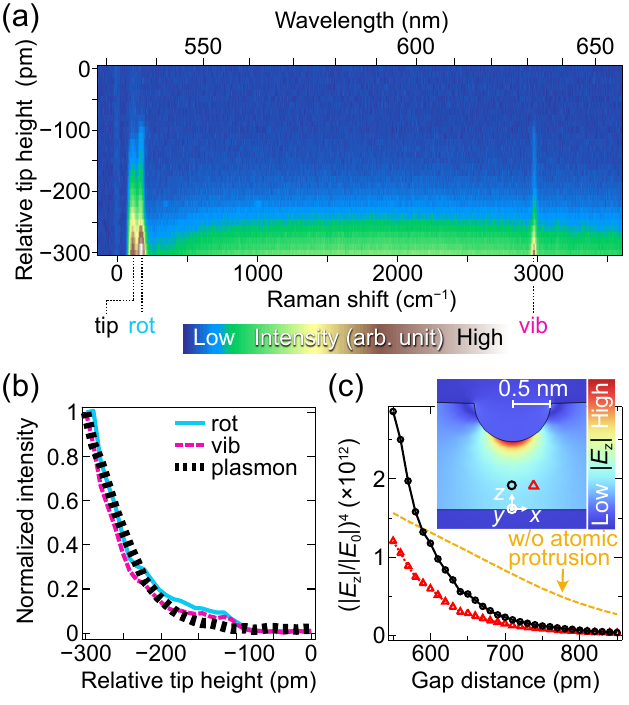}
\caption{\label{fig2}
(a) Waterfall plot for the TERS spectra during the tip approach to D$_2$/Ag(111).
The origin of the vertical axis corresponds to a tip height 100 pm above the STM set point ($V_\mathrm{s}$ = 10 mV, $I_\mathrm{t}$ = 0.1 nA).
(b) Tip-height dependence of the Raman intensities, corresponding to the vertical line profiles of ``rot,'' ``vib,'' and the plasmonic background (2000 cm$^{-1}$).
The intensity is normalized for each profile.
(c) Gap-distance dependence of the field enhancement factor simulated by the finite element method.
The STM junction with a plasmonic picocavity is modeled using an Ag tip (30-nm radius of curvature) with an atomic-scale protrusion (0.5-nm radius) and a flat Ag sample plate (Fig.~S9).
The vertical electric field $E_z$ at $\lambda_\mathrm{exc} = 532$ nm is sampled and normalized with the incident field $E_0$.
The inset shows the $|E_z|$ distribution at a tip--sample gap distance of 0.9 nm.
The black curve represents the field enhancement factor $(|E_z|/|E_0|)^4$ sampling at the location of the D$_2$ just below the tip (circle marker in the inset), while the red curve is that at the nearest neighboring D$_2$ (triangle).
The orange curve shows the field enhancement at the same sampling point as the black curve but using an Ag tip without the atomic-scale protrusion.
}
\end{figure}

The TERS intensity is strongly correlated with the localized surface plasmonic resonance (LSPR) of the tip--sample junction.
In general, two types of enhancement mechanisms contribute to TERS \cite{itoh2023toward,chen2023interpreting}: electromagnetic enhancement through LSPR  and chemical enhancement originating from charge transfer between the target molecule and the tip or surface.
In physisorption systems, the electromagnetic enhancement is expected to be dominant.

To elucidate the role of electromagnetic enhancement, it is crucial to investigate the gap-distance dependence of TERS signals. 
Figure~\ref{fig2}(a) shows the waterfall plot of the gap-distance-dependent TERS spectra recorded for D$_2$/Ag(111).
When the tip approaches the surface, the TERS intensities of the rotational and vibrational modes increase apparently exponentially [Fig.~\ref{fig2}(b)], in agreement with the previously reported analytical model simplifying a plasmonic tip as a point dipole \cite{pettinger2009tip}.
A broad background with a maximum at $\sim$590 nm is attributed to electronic Raman scattering which reflects the spectral response of the LSPR in the junction \cite{kamimura2022surface}.
The gap-distance-dependent intensities of both rotational and vibrational modes closely follow the plasmonic feature [Fig.~\ref{fig2}(b)], indicating the dominant contribution of the gap-mode plasmon to the TERS enhancement.
Additionally, we observed that changes in both the plasmon background and the molecular Raman peak occurred by an accidental modification of the tip-apex structure (Fig.~S11), implying that the TERS enhancement is susceptible to the atomic-scale structure of the tip apex, which generates a plasmonic picocavity.

The gap-distance dependence of the TERS intensities is well reproduced by simulating the electronic field distribution in the Ag-tip--Ag-surface junction using the finite element method (see SM Sec.~I-E \cite{SI}).
To model the plasmonic picocavity, we assume an atomic-scale protrusion (0.5-nm radius of curvature) attached to a blunt tip body (Fig.~S9).
The black curve in Fig.~\ref{fig2}(c) shows the field enhancement at the expected position of a D$_2$ molecule located beneath the Ag tip, indicating a steep increase as the tip approaches.
We confirmed the critical role of this subnanometric structure in TERS; as shown by the orange curve in Fig.~\ref{fig2}(c), the tip without the atomic-scale protrusion exhibits a more moderate increase of the field enhancement as the tip approaches, which does not reproduce the experimental data [Fig.~\ref{fig2}(b)].
The red curve in Fig.~\ref{fig2}(c) shows the field enhancement at a possible position of the nearest neighboring molecule [Fig.~\ref{fig2}(c), inset], indicating much lower enhancement.
Therefore, TERS can achieve ``nearly'' single-molecule sensitivity, even for physisorption systems. 
The contribution of LSPR to the TERS intensity was further confirmed by our excitation-wavelength dependent measurements (SM Sec.~II-B \cite{SI}); a strong TERS peak was observed only when the scattering light wavelength matches with the resonance range of the LSPR (Fig.~S10).

\begin{figure}[ht]
\includegraphics[width=0.48\textwidth]{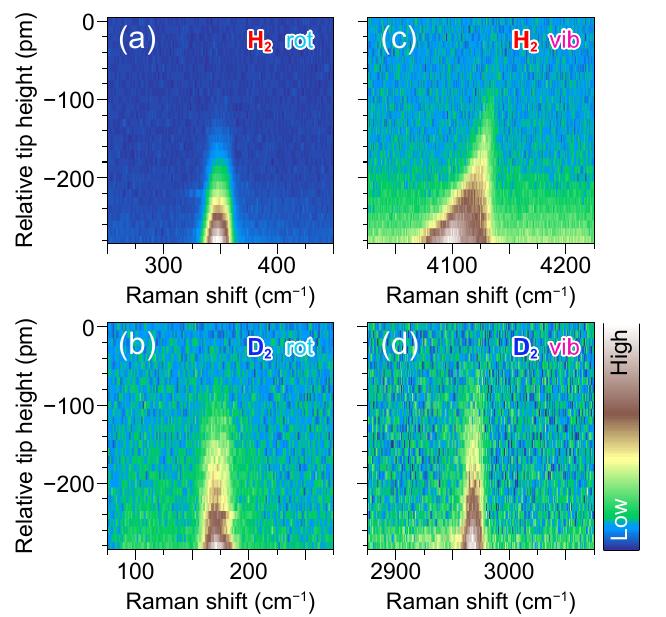}
\caption{\label{fig3}
(a--d) Waterfall plots of the high-resolution TERS spectra with different tip heights: (a) H$_2$ rot, (b) H$_2$ vib, (c) D$_2$ rot, and (d) D$_2$ vib. 
The origin of the tip height corresponds to a tip height 30 and 70 pm higher than the set point ($V_\mathrm{s} = 10$ mV, $I_\mathrm{t} = 0.1$ nA) for H$_2$ and D$_2$, respectively.
}
\end{figure}
\begin{figure}[ht]
\includegraphics[width=0.46\textwidth]{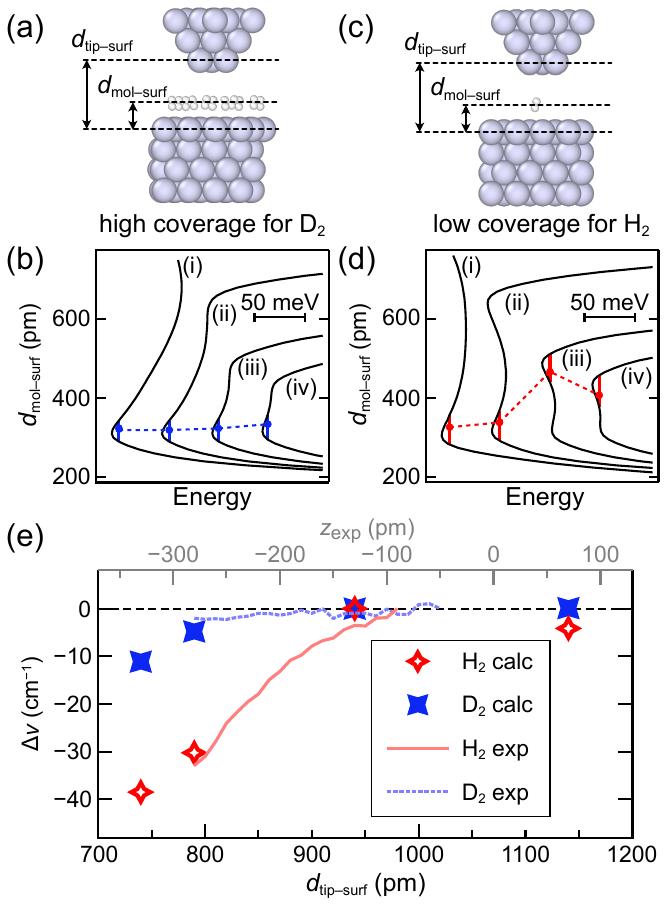}
\caption{\label{fig4}
(a) Simulated atomic structure of the junction at a tip--surface distance $d_\mathrm{tip-surf}$ of 0.79 nm at a high coverage (0.69 ML).
The light blue and white spheres represent Ag and H atoms, respectively.
(b) Energy profiles along $d_\mathrm{mol-surf}$ at the high coverage shown in (a).
Curves (i)--(iv) indicate profiles obtained at $d_\mathrm{tip-surf}$ = 1.14, 0.94, 0.79, and 0.74 nm, respectively.
For clarity, curves (ii)--(iV) are offset by 50, 100, and 150 meV, respectively, relative to (i) in the horizontal axis.
The bullet on each potential indicates the average D$_2$ position, considering ZPE contributions (Table S3).
The vertical lines indicates the ZPE levels and the dashed lines connecting the bullets serve only as a guide to the eye.
(c) Same as (a) but at a lower coverage (0.06 ML).
(d) Same as (b), but at the lower coverage shown in (c).
The bullet on each potential indicates the average H$_2$ position, considering ZPE contributions (Table S2).
(e) Shifts of the intramolecular stretch frequencies $\Delta \nu$ as a function of $d_\mathrm{tip-surf}$.
The empty red and filled blue bullets represent the calculated values of low-coverage H$_2$ and high-coverage D$_2$, respectively.
The frequencies at $d_\mathrm{tip-surf}$ = 0.94 nm are used for the origin of the shifts.
The solid red and dashed blue lines represent the experimental $\Delta \nu$ values for H$_2$ vib [Fig.~\ref{fig3}(c)] and D$_2$ vib [(d)], respectively, as a function of the relative tip height $z_\mathrm{exp}$ (see SM Sec.~II-C \cite{SI} for details).
The minimum tip height, $z_\mathrm{exp} = -280$ pm, is aligned with $d_\mathrm{tip-surf}$ = 0.74 nm.
}
\end{figure}
\clearpage

TERS reveals the sensitivity of the rotational and vibrational states to the interactions between a target molecule and its local environment.
Figures~\ref{fig3}(a)--\ref{fig3}(d) show the gap-distance dependence of high-resolution TERS spectra focusing on the rotational and vibrational modes of H$_2$ and D$_2$.
Notably, only the vibrational frequency of H$_2$ exhibits a significant redshift as the gap distance decreases [Fig.~\ref{fig3}(c)]; 
H$_2$ shows a peak shift $\Delta \nu$ of $-$33 cm$^{-1}$ with a 200 pm displacement of the tip, whereas $\Delta \nu$ is merely $-2$ cm$^{-1}$ for D$_2$.
Additionally, the peak width of the H$_2$ vibrational mode also increases as the gap distance decreases (see SM Sec.~II-C \cite{SI} for details).
In contrast, the peak positions of the rotational modes for both H$_2$ [Fig.~\ref{fig3}(a)] and D$_2$ [\ref{fig3}(b)] are almost unchanged, consistent with our calculations showing that the interatomic distance is independent of the tip height, as described above.

The isotope dependence of the redshift of the vibrational band far exceeds the scale of typical isotope effects.  
To understand this anomalous behavior, 
we calculated the potential energy profile $V(d_{\rm{mol-surf}}, d_{\rm{tip-surf}})$ along the rigid vertical displacement of the center of mass of one molecule between the surface and the tip apex $d_{\rm{mol-surf}}$ [Fig.~\ref{fig4}(a)], while varying the tip--sample distance $d_{\rm{tip-surf}}$ [see profiles (i)--(iv) at different $d_{\rm{tip-surf}}$ in Fig.~\ref{fig4}(b)].
At each $d_{\rm{mol-surf}}$ position, we calculated the H$_2$ and D$_2$ harmonic intramolecular vibrational frequency $\nu_{\mathrm{harm}}$ (see SM Sec.~I-C \cite{SI}).
The calculations indicate a pronounced coupling of the stretch coordinate $r_{\rm{stretch}}$ with the molecular translation along $d_{\rm{mol-surf}}$, evidenced by a steep variation of $\nu_{\mathrm{harm}}$ along this coordinate (Fig.~S7).
Furthermore, this variation changes depending on the tip--sample distance, leading us to define $\nu_{\mathrm{harm}}(d_{\rm{mol-surf}}, d_{\rm{tip-surf}})$. 
As detailed in SM Sec.~I-C \cite{SI}, for each $d_{\rm{tip-surf}}$, we constructed a model 2D Schr{\"o}dinger equation in the coupled coordinates $d_{\rm{mol-surf}}$ and $r_{\rm{stretch}}$, parameterized by $V(d_{\rm{mol-surf}}, d_{\rm{tip-surf}})$ and $\nu_\mathrm{harm}(d_{\rm{mol-surf}}, d_{\rm{tip-surf}})$. 
The solution of this model gives us the expected intramolecular stretch vibrational transition energy $\nu$ for each isotope, fully accounting for the anharmonic coupling between these coordinates. 
We concluded that if both H$_2$ and D$_2$ probe the same potential energy profile $V(d_{\rm{mol-surf}}, d_{\rm{tip-surf}})$ [Fig.~\ref{fig4}(b)],
for reasonable position-dependent stretch frequencies $\nu_\mathrm{harm}(d_{\rm{mol-surf}}, d_{\rm{tip-surf}})$ the calculated peak shifts $\Delta \nu$ are in a ratio of at most \mbox{2:1} for \mbox{H$_2$:D$_2$} (Tables~S2 and S3), in disagreement with the measured \mbox{16:1} ratio (Fig.~\ref{fig3}).

We demonstrate that this discrepancy can be resolved by accounting for density differences between H$_2$ and D$_2$.
At lower temperatures, the solids formed by H$_2$ and D$_2$ have different molar volumes due to nuclear quantum effects [zero-point energy (ZPE) swelling \cite{MarklandManoJCP2008}; see also SM Sec.~I-B \cite{SI}]. As a result, solid D$_2$ is more than twice as dense as solid H$_2$ at 4.2~K~\cite{MegawSimon2936}.  
Because the interaction potential between the molecules and the surface exhibits almost no corrugation \cite{kunisada2015hindered}, we expect that H$_2$ adsorbs on Ag(111) at 10 K more sparsely than D$_2$.
To explore a limiting low-density regime, we repeated the calculations discussed in the previous paragraph for a coverage of 0.06 ML of H$_2$ [Fig.~\ref{fig4}(c)]. 
As shown in Fig.~\ref{fig4}(d), the reduced intermolecular interactions at the surface cause the minimum of $V(d_{\rm{mol-surf}}, d_{\rm{tip-surf}})$ along $d_{\rm{mol-surf}}$ to lie closer to the tip instead of closer to the surface, as $d_{\rm{tip-surf}}$ decreases [(i) to (iv)]. 
Consequently, the average position of the molecule center of mass [bullet points in Figs.~\ref{fig4}(b) and (d)] also shifts toward the tip as $d_{\rm{tip-surf}}$ decreases [dotted line in Fig.~\ref{fig4}(d)].
Computing $\nu$ in this scenario with the 2D quantum model of $d_{\rm{mol-surf}}$ and $r_{\rm{stretch}}$, we find a larger variation with $d_{\rm{tip-surf}}$ than in the higher-density case (Table~S2). 
As shown in Fig.~\ref{fig4}(e), considering the higher-density scenario for D$_2$ (the filled bullets) and the lower-density scenario for H$_2$ (the empty bullets) can account for the large isotope effect in the redshift of the stretch transition as the tip approaches.
Figure~\ref{fig4}(e) also shows the experimental $\Delta \nu$ values from the plots in Fig.~\ref{fig3} (the solid and dotted curves for H$_2$ and D$_2$, respectively; see also Fig.~S12).
Although the exact gap distance in STM experiment is hard to determine (see SM Sec.~II-C \cite{SI}), the overlap of the experimental curves with the calculated plots in Fig.~\ref{fig4}(e) shows good agreement.
The remaining numerical discrepancies can be attributed to the limitation of the DFT level of theory (see SM Sec.~I-B \cite{SI}), the exact tip-apex geometry, and to the uncertainty in determining the exact local surface coverage. 
Nevertheless, our results highlight the crucial role of intermolecular interactions in accurately interpreting vibrational spectroscopy of hydrogen isotopes in nanojunctions, where nuclear quantum effects have a significant impact \cite{williams2002raman}.

In summary, with LT-TERS, we observed the rotational and vibrational transitions of local H$_2$ and D$_2$ molecules through the interaction with the picocavity field.
Our results demonstrate that LT-TERS is applicable to weakly adsorbed molecules, offering deeper insights into the local structures, reactions, and dynamics of adsorbates, such as chemical reactivity at active sites \cite{cai2023probing} and surface diffusion (spillover) in catalytic processes \cite{yang2024insights}, at the single-molecule level.

\begin{acknowledgments}
\emph{Acknowledgments---}We thank Adnan Hammud for the Ag-tip fabrication. 
We thank Jun Yoshinobu and Heiko Appel for helpful discussions.
T.S.~acknowledges the support of JST FOREST Grant No. JPMJFR221U.
T.K.~acknowledges the support of JST FOREST Grant No.~JPMJFR201J.
\end{acknowledgments}





%



\newpage

\setcounter{equation}{0}
\setcounter{figure}{0}
\setcounter{table}{0}
\setcounter{page}{1}
\makeatletter
\renewcommand{\thepage}{S\arabic{page}}
\renewcommand{\theequation}{S\arabic{equation}}
\renewcommand{\thefigure}{S\arabic{figure}}
\renewcommand{\thetable}{S\arabic{table}} 

\newcommand{\dH}{\ensuremath{d_{\mathrm{H}_2}^\mathrm{qc}}}
\newcommand{\dD}{\ensuremath{d_{\mathrm{D}_2}^\mathrm{qc}}}

\section*{Supplemental Material: P\lowercase{icocavity}-E\lowercase{nhanced} R\lowercase{aman} S\lowercase{pectroscopy of} P\lowercase{hysisorbed} H$_2$ \lowercase{and} D$_2$ M\lowercase{olecules}}



\section{I. Methods}

\subsection{A. STM-TERS Experiments}

\begin{figure*}[bth]
\includegraphics[width=14cm]{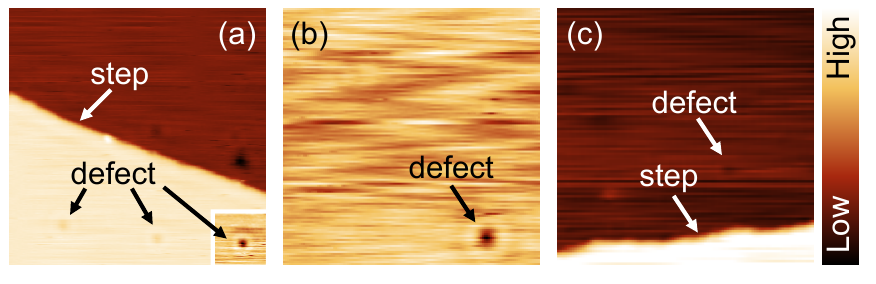}
\caption{\label{figS1}
(a)--(c) Typical STM images of a clean Ag(111) surface, H$_2$/Ag(111), and D$_2$/Ag(111), respectively [$V_\mathrm{s}$ = 0.1 V, $I_\mathrm{t}$ = 0.1 nA, $T$ = 10 K, exposure: H$_2$ 50 L for (b) and D$_2$ 220 L for (c), scan size: 40 $\times$ 40 nm$^2$ for (a) and (c), 24 $\times$ 24 nm$^2$ for (b)].
The inset of (a) shows the STM image of a defect on the clean Ag(111) surface with another color scale.
The TERS spectra of H$_2$/Ag(111) and D$_2$/Ag(111) shown in Fig.~1b of the main text were obtained with the same sample and tip as (b) and (c), respectively.
The circular depressions observed in (b) and (c) are ascribed to subsurface point defects, which also appear on a clean surface as shown in (a) and its inset STM image. 
Although the appearance of subnanoscale structures such as detects and single Ag atomic steps confirms high spatial resolution imaging, no static hydrogen molecule was observed under the condition used.
}
\end{figure*}

\begin{figure*}[tbh]
\includegraphics[width=11cm]{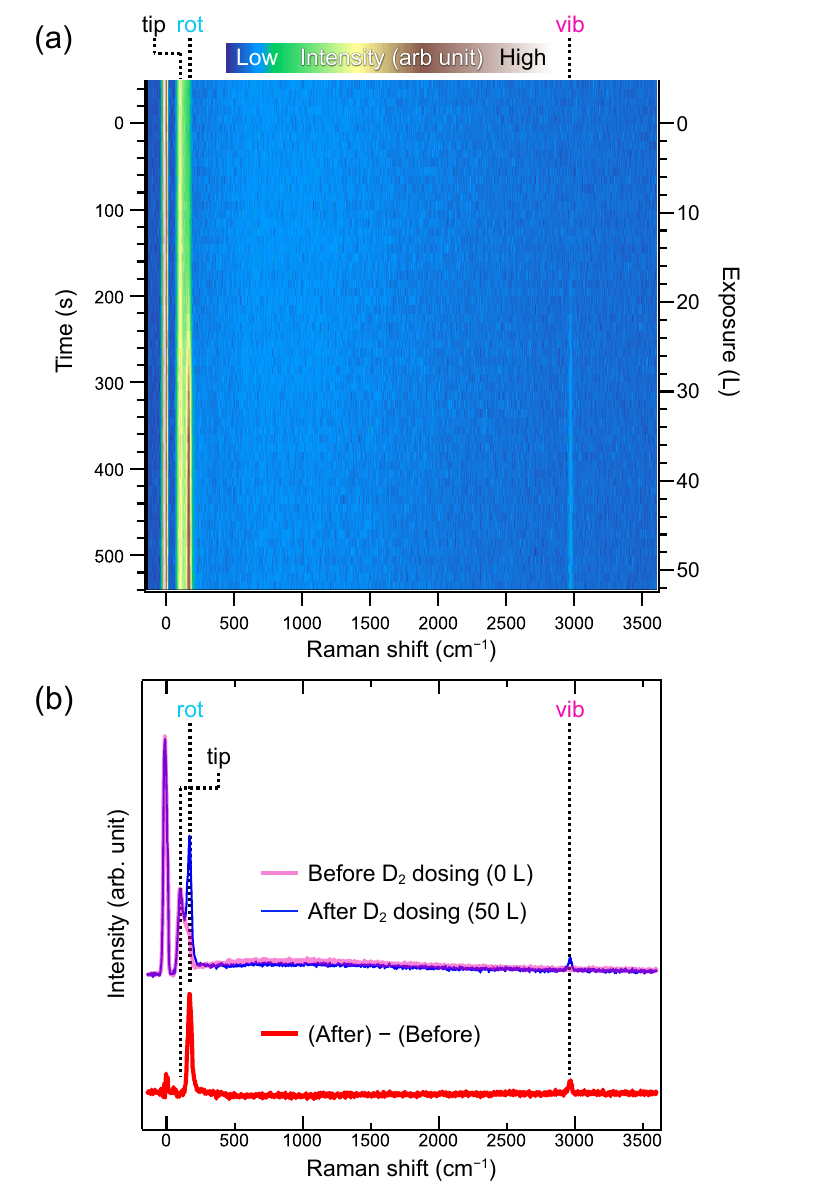}
\caption{\label{figS2}
(a) Waterfall plot of TERS spectra of Ag(111) during D$_2$ gas exposure ($V_\mathrm{s}$ = 0.01 V, $I_\mathrm{t}$ = 1 nA, $T$ = 10 K, $\lambda_\mathrm{exc} = 532$ nm, accumulation: 10 s/spectrum).
The point at which the gas dosing started is set as the origin of the time axis.
The ``rot'' and ``vib'' modes of \textit{ortho}-D$_2$ are indicated on top of the plot.
The structure at $\sim$100 cm$^{-1}$ that was constantly observed before the gas dosing is ascribed to the Ag phonon mode (labeled as ``tip''), as described in the main text.
(b) TERS spectra of the sample before (bold magenta curve; 0 L) and after (narrow blue curve; 50 L) D$_2$ gas exposure.
The red curve represents the post-dose spectrum subtracted from the pre-dose spectrum, where the ``tip'' structure is canceled out and only the molecular peaks appear.
}
\end{figure*}

The scanning tunneling microscopy (STM) and tip-enhanced Raman spectroscopy (TERS) experiments were performed in an ultrahigh vacuum (UHV) chamber (modified UNISOKU USM-1400) at a sample temperature $T$ of 10 K.
A single-crystalline Ag(111) surface (MaTeck) was cleaned by Ar-sputtering-and-annealing cycles in the chamber.
H$_2$ ($\geq$99.999\%, Westfalen AG) or D$_2$ (99.7\%, Air Liquide) gas was introduced into the STM unit cooled at 10 K via a variable leak valve.
The gas exposure (Langmuir; L) is estimated by $Pt/n$, where $P$ [$10^{-6}$ Torr], $t$ [s], and $n$ denote the displayed pressure value of the gauge (corrected by N$_2$), the exposure duration, and the correction factor for the gauge, respectively.
A Bayard-Alpert ionization gauge is located outside the cryogen-cooled STM unit in the chamber and we used $n = 0.384$ for H$_2$ and $n = 0.388$ for D$_2$ \cite{summers1969empirical}. 
We used a chemically-etched Ag tip that was further sharpened by focused-ion-beam milling as described elsewhere \cite{liu2019resolving}.
The bias voltage $V_\mathrm{s}$ was applied to the sample while the tip was grounded.
The STM images were obtained in the constant current mode at tunneling current $I_\mathrm{t}$ of 50 pA.
The cleanness of the Ag(111) surface and tip was confirmed by STM images [Fig.~\ref{figS1}(a)].
We also confirmed that a TERS spectrum of the Ag--Ag junction before hydrogen dosing had no molecule-derived peaks (Fig.~\ref{figS2}, at 0 L). 

In the STM junction, $p$-polarized visible laser (532-nm solid-state or 633-nm HeNe) was directed through a fused silica window and an Ag-coated parabolic mirror. 
The parabolic mirror is mounted on a five-axis piezoelectric-motor stage in the low-temperature (LT) STM unit and the laser is focused at the tip apex by adjusting the position and angle of the parabolic mirror. 
The scattered light collected by the parabolic mirror was directed to a spectrometer (AndorShamrock 303i) outside the UHV chamber via a beam splitter and a longpass filter, each selected according to the excitation wavelength $\lambda_\mathrm{exc}$.
The nominal incident laser power is 5.9 mW at $\lambda_\mathrm{exc}$ = 532 nm and 9.7 mW at $\lambda_\mathrm{exc}$ = 633 nm.
For the TERS and STM-induced luminescence (STML) spectra, we used the gratings 150 (for Fig.~\ref{figS3}), 600 (Figs.~1, 2, and \ref{figS2}), and 1200 line/mm (Fig.~3 and \ref{figS11}).
A lower-density grating allows the acquisition of spectra over a wider wavelength range, as shown in Fig.~\ref{figS3} (see Sec.~II-B), but resulting in broad and asymmetric Raman peaks.
Therefore, for the high-frequency-resolution measurements shown in Fig.~3, the densest grating was used.


\clearpage

\subsection{B. DFT Calculations and Molecular Dynamics%
\label{sec:dft}}

\begin{figure}[htb]
    \centering
    \includegraphics[width=0.7\textwidth]{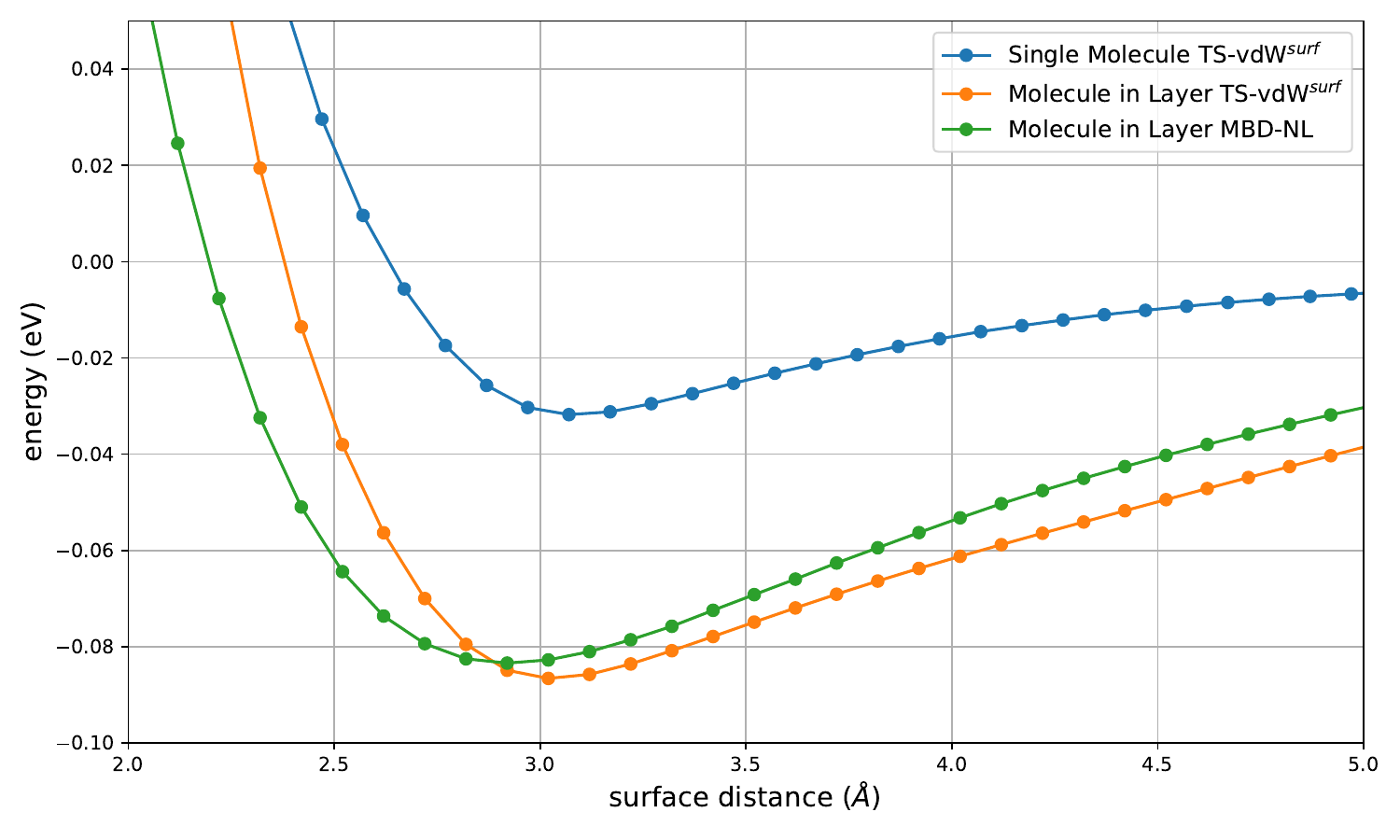}
    \caption{Binding energies of H$_2$ to the Ag(111) surface calculated as the following difference in total energies: 
     $E_\mathrm{b}(d_\mathrm{mol-surf}) = E_{\mathrm{H}_2/\mathrm{Ag}(111)}(d_\mathrm{mol-surf}) - E_{\mathrm{H}_2/\mathrm{Ag}(111)}(10~\mathrm{\AA})$. In other words, the zero of energy is taken to be the total energy value $E_{\mathrm{H}_2/\mathrm{Ag}(111)}$ when H$_2$ is 10~{\AA} away from the surface.
    The blue curve was calculated with a a coverage of $\sim$0.06 ML (1 molecule per 4 $\times$ 4  surface cell), while the orange and green curves were calculated for a coverage of $\sim$0.69 ML (11 molecules per surface cell).
    The PBE functional with TS-vdW$^\mathrm{surf}$ dispersion interactions were used for the blue and orange curves and the PBE functional with the MBD-NL dispersion correction was used for the green curve (see text).
    }
    \label{fig:binding-energies}
\end{figure}

\begin{figure}[ht]
    \centering
    \includegraphics[width=0.3\textwidth]{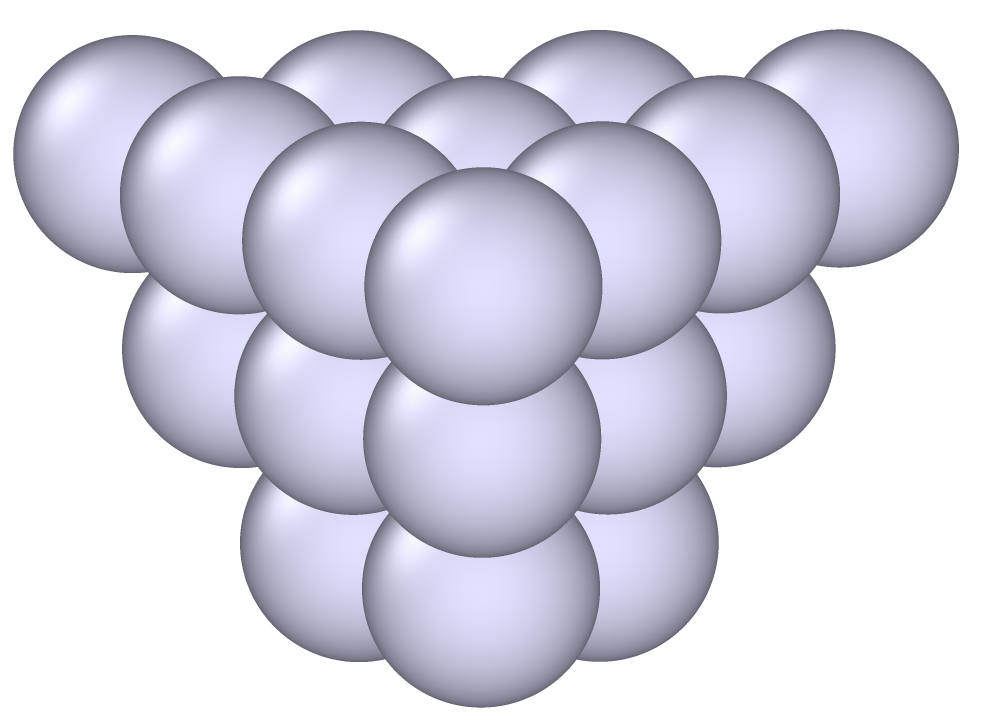}
    \caption{Geometry used for the model tip structure (Ag$_{19}$ cluster).\label{fig:tip}}
\end{figure}

Density functional theory (DFT) calculations were performed with the FHI-aims program package \cite{blum+cpc2009, gavi+roadmap2023}.
We used intermediate defaults for basis sets, radial grids and cutoff potential parameters and light defaults for the Hartree multipole expansion and angular grids.
This combination proved efficient and reliable against benchmarks to tight settings.
We simulated an orthogonal $4 \times 4$ Ag(111) surface cell oriented perpendicular to the $z$ axis, containing four layers.
The cell parameters used were $a = 11.554$ {\AA}, $b = 10.007$ {\AA} and $c = 65.000$ {\AA}.
We employed a $2 \times 2 \times 1$ $k$-point grid and a dipole correction was employed to maximally decouple periodic images along the $z$ direction.
We used the Perdew-Burke-Ernzerhof (PBE) exchange-correlation functional and included van der Waals (vdW) interactions through the pairwise screened TS-vdW$^\mathrm{surf}$ method of Ref.~\citep{ruiz2012prl} in all simulations reported in the main text.
Figure~\ref{fig:binding-energies} shows a comparison of binding energies with the many-body dispersion correction termed MBD-NL in Ref.~\citep{hermann2020prl}, indicating a weaker binding energy by only 5 meV and a binding distance 0.1 {\AA} closer to the surface.
Structures were optimized by allowing only hydrogen molecules and the top layer of Ag atoms of the surface to relax.
All the other atoms were kept fixed at their bulk (or tip) positions.
When we performed simulations including a model tip structure, this was modeled by an Ag$_{19}$ cluster at varying distances from the surface. 
The geometry of the tip is shown in Fig.~\ref{fig:tip}.
Although this tip size would not be sufficient to capture a proper plasmonic response of the tip, the tip structure can capture the local interactions with molecules close to the tip apex \cite{litman2023jpcl,cirera2022charge,liu2023inelastic}.

\clearpage

We inspected binding sites and binding energies of a single adsorbed H$_2$ molecule, finding all values in good agreement with previous studies \cite{kunisada2015hindered}.
The binding energy was defined as the difference between the total energy when the molecule was 10 {\AA} away from the surface and the minimum total energy of the adsorbed molecule.
The blue curve in Fig.~\ref{fig:binding-energies} shows the binding curve of a single H$_2$ molecule adsorbed on the face-centered-cubic (fcc) hollow site of the Ag(111) surface, indicating an equilibrium H$_2$--Ag(111) bonding length $d_\mathrm{opt}$ of 3.09 {\AA} and a binding energy of 32 meV.
The model we used in most simulations contained 11 molecules in the unit cell (see rationale below).
For this molecular concentration, we evaluated the intermolecular interaction by comparing the binding curves (orange curve in Fig.~\ref{fig:binding-energies}) with that of the single-molecule case (blue);
the molecule--molecule interaction in our models amounts to 55 meV/molecule.

Performing vibrational analysis on the optimized single molecule adsorbed on the surface showed that negative frequencies related to rotational and translational modes of the molecule were always present.
This led us to consider higher coverages to model local stable clustering of the molecules on the surface.
We started by considering a full monolayer (1 ML, i.e., 16 molecules on the $4 \times 4$ surface cell), where all fcc-hollow sites of the unit cell were occupied by H$_2$.
With this setup we performed Born-Oppenheimer molecular dynamics (BOMD) and path-integral molecular dynamics (PIMD) at LT ($T$ = 20 K) with the i-PI code \cite{kapil+cpc2019} connected to FHI-aims.
PIMD simulations used 12 beads in conjunction with a normal-mode colored-noise thermostat with 8 auxiliary variables and $\hbar \omega / kT = 500$.
This number of beads is not fully converged at this temperature even with the colored-noise quantum thermostats, but the purpose of these simulations was exploratory. 

While the full monolayer was stable during more than 10 ps on the classical trajectories, in the quantum trajectory 5 molecules readily desorbed (within the first picosecond) from the substrate, leaving only 11 molecules adsorbed, corresponding to a $\sim 0.69$ ML coverage. This is the coverage adopted in other simulations in this Letter as the high-density model. 
For the quantum case, we validated this observation by simulating trajectories with a parameterized Lennard-Jones potential representing the Ag(111) surface.
We could calculate the radius of gyration of H atoms from the PIMD simulations, belonging to the 3 molecules closest to the tip at any point in the simulation.
For the closest tip distance considered ($d_\mathrm{tip-surf}$ = 7.4 {\AA}), these were 0.33 {\AA} along $x$ and $y$ and 0.30 {\AA} along $z$ ($T$ = 20 K), indicating a small anisotropy.
Converting these numbers to an effective thermal de Broglie wavelength $\Lambda$ and applying a inverse-square-root scaling for the temperature and mass, we would obtain $\Lambda = 2.3$ {\AA} for H$_2$ and $1.5$ {\AA} for D$_2$ at 10 K.
We note that, depending on the density of H$_2$ molecules on the surface, $\Lambda$ for H$_2$ is about the length of the inter-particle spacing at 10 K and signatures of bosonic exchange could begin to be observed.
These features are disregarded in the current simulations. 
Further, we note that even though these simulations provide a clear indication that quantum mechanical effects affect the density of the molecular coverage of the surface, a complete determination would require much longer simulations, which would take years to complete or require the development of a tailored machine-learning potential that surpasses the scope of this work.

\begin{figure}[htb]
    \centering
    \includegraphics[width=0.5\textwidth]{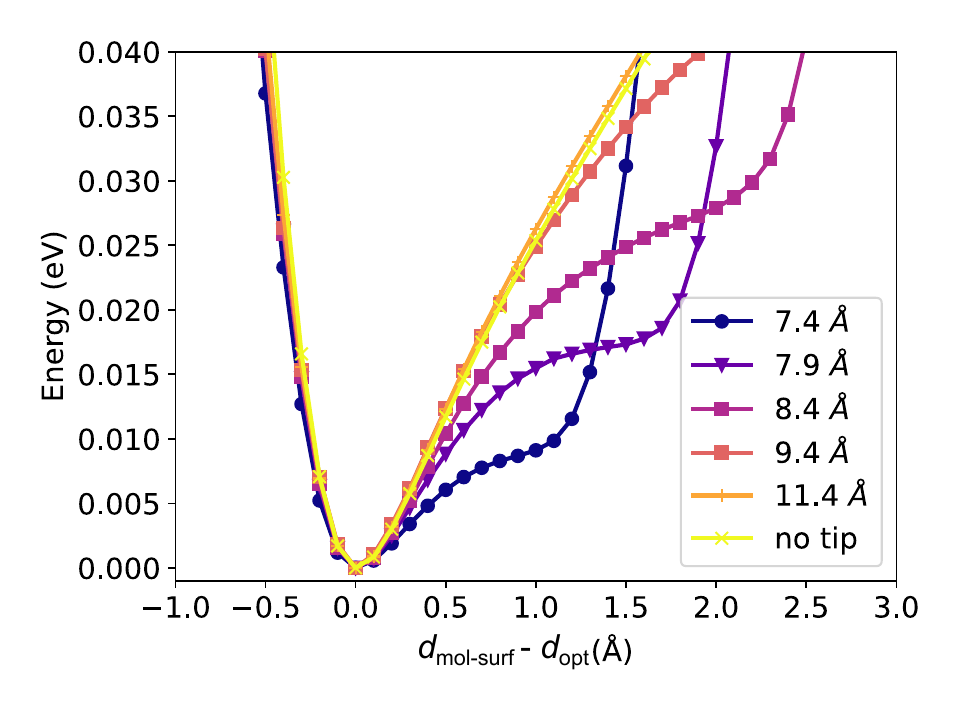}
    \caption{Energy profiles of an H$_2$ molecule with varying distance to the Ag(111) surface without (yellow curve) and with an Ag tip at different tip--surface distances (the other curves).
    All results obtained with the PBE+TS-vdW$^\mathrm{surf}$ exchange-correlation functional.
    For visual clarity, each curve was shifted so that the equilibrium H$_2$--Ag(111) bonding distance and the bonding energy at each tip height is at the origin of the horizontal and vertical axes, respectively.}
    \label{fig:energy-profile}
\end{figure}

\begin{figure}[htb]
    \centering
    \includegraphics[width=0.9\textwidth]{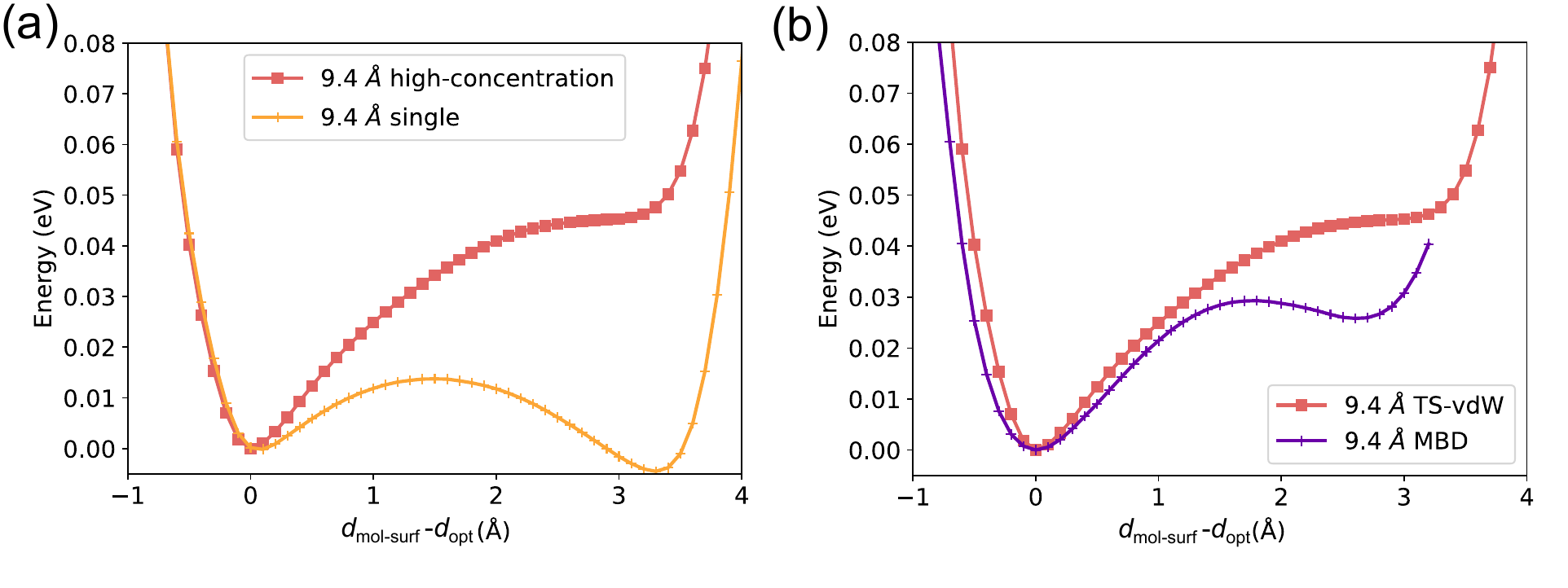}
    \caption{(a) Energy profiles of an H$_2$ molecule with varying distance to the Ag(111) surface at a tip--surface distance of 9.4 {\AA}.
    The red and orange curves indicate the profiles with the higher concentration (11 molecules in the cell) and the lowest concentration (1 molecule in the cell), respectively.
    The origins of the axes are defined in the same manner as in Fig.~\ref{fig:energy-profile}. (b) Comparison of the high-concentration curve with different vdW corrections (TS-vdW$^\mathrm{surf}$~\cite{ruiz2012prl} and MBD-NL~\cite{hermann2020prl}).
    }
    \label{fig:comparison-potential}
\end{figure}

Figure~\ref{fig:energy-profile} shows the energy profiles experienced by an H$_2$ molecule when varying the distance of its center of mass with respect  to the Ag(111) surface, at different tip-surface distances, and at 0.69 ML coverage.
We only varied the position of the molecule closest to the tip apex, while the others were kept at equilibrium.
The profile shapes are similar to an asymmetric double-well potential at closer tip--surface distances, while they assume a shape closer to a Morse potential at larger distances, which asymptotically approaches the profile of an H$_2$ molecule on Ag(111) without the Ag tip (yellow curve in Fig.~\ref{fig:energy-profile}).
The asymmetry of the double-well is strongly impacted by the presence of other molecules on the surface (molecule--molecule interactions).
Figure~\ref{fig:comparison-potential}(a) shows a comparison with the limiting case of a single molecule in the unit cell (0.06 ML coverage) at a tip--surface distance of 9.4 {\AA}.
In that case (orange curve), if the molecule can overcome the barrier, it will have a higher probability of being found closer to the tip, which was predicted in a Au-tip--single-H$_2$--Au-surface junction \cite{sugimoto2017inelastic}. As we discuss in the main text, this effect is essential to explain the isotope-dependent redshift observed experimentally.

Figure~\ref{fig:comparison-potential}(b) shows a comparison of the 0.69 ML coverage with the tip fixed at $d_{\mathrm{tip-surf}}= 9.4$~\AA~and different vdW dispersion corrections, added to the PBE functional. It is clear that details of the vdW dispersion impact the shape of the potential. In this case, the many-body vdW dispersion decreases the distance between the two minima of the asymmetric double-well. This change would impact the quantitative value of the tip--surface distance at which H$_2$ molecules would prefer to sit closer to the tip, at the low-density limit. We do not expect qualitative changes on the effects we discuss in the main text. 


\subsection{C. Model 2D Schr\"{odinger} Equation}

To model the coupling between the intramolecular vibration and the restricted translation of an \ce{H2}/\ce{D2} molecule between the substrate and the tip, we constructed a series of two-dimensional (2D) potentials, parametrized by the tip--surface distance $d_{\text{tip--surf}}$,
\begin{equation}
    \label{eq:model-pes}
    U_{\text{2D}}(d_{\text{mol--surf}}, r_{\text{stretch}}; d_{\text{tip--surf}}) = V(d_{\text{mol--surf}}, d_{\text{tip--surf}}) + \frac{\mu r_{\text{stretch}}^2 \omega^2(d_{\text{mol--surf}}, d_{\text{tip--surf}})}{2},
\end{equation}
where $V(d_{\text{mol--surf}}, d_{\text{tip--surf}})$ is the potential energy of a molecule with its center of mass fixed at a distance $d_{\text{mol--surf}}$ above the substrate surface and with all other coordinates kept at their equilibrium values, as described at the end of Sec.~\ref{sec:dft}. The vibrational motion along the intramolecular stretching coordinate $r_{\mathrm{stretch}}$ is parametrized by the reduced mass $\mu$ and angular frequency $\omega(d_{\text{mol--surf}}, d_{\text{tip--surf}})$. We set the mass of a hydrogen atom  to $m_{\ce{H}} = 1836.1527\,m_{\text{e}}$ and use $
    m_{\ce{D}} = 2 m_{\ce{H}}, \, \mu_{\ce{H2}} = m_{\ce{H}}/2, \, \text{and} \ \mu_{\ce{D2}} = m_{\ce{D}}/2$, with $m_{\text{e}}$ denoting the mass of the electron.
    
\begin{figure*}[htb]
    \includegraphics{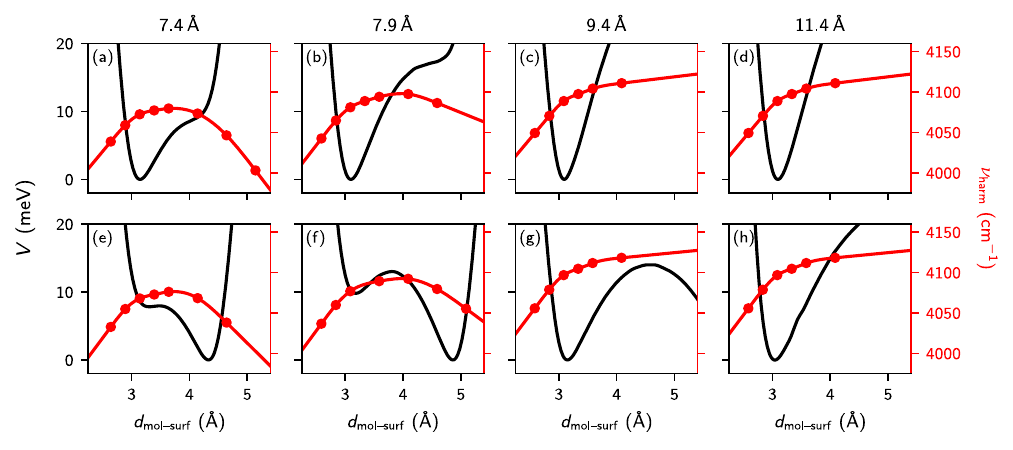}
    \caption{Translational potential energy profiles $V(d_{\text{mol--surf}}, d_{\text{tip--surf}})$ and 
    \ce{H2} vibrational wavenumbers $\nu_{\text{harm}}(d_{\text{mol--surf}}, d_{\text{tip--surf}})$ corresponding to the frequencies in Eq.~\eqref{eq:sam}, shown with black and red lines, respectively.
    Panels (a--d) show the parameters for 0.69~ML coverage, and panels (e--f) show the parameters for 0.06~ML coverage.
    The headings above the panels show the values of $d_{\text{tip--surf}}$.
    The translational energies were computed on a dense grid, whereas the frequencies were calculated at a select few points, shown with red circles. We use cubic spline interpolation to construct smooth curves and linear extrapolation to extend the curves beyond the grid boundaries. Values of $\nu_{\text{harm}}$ at 
    \mbox{$d_{\text{tip--surf}} = 11.4\,\text{\AA}$} are copied over from \mbox{$d_{\text{tip--surf}} = 9.4\,\text{\AA}$}, since point calculations confirm that harmonic stretch frequencies are essentially independent of $d_{\text{tip--surf}}$ over the relevant range of $d_{\text{mol--surf}}$ at these tip--surface separations.
    \label{fig:model-pes}}
\end{figure*}

The vibrational frequencies are computed from numerical (finite-difference) second derivatives of the DFT energies with respect to the vibrational coordinate,
\begin{equation}
    \label{eq:sam}
    \omega_{\ce{X2}} = \sigma \times \sqrt{\frac{1}{\mu_{\ce{X2}}} \pdv[2]{V}{r_{\text{stretch}\mkern-12mu}} },
\end{equation}
where $\ce{X} = \ce{H} \text{ or } \ce{D}$ and $\sigma = 0.96364$ is a uniform scaling factor that accounts for the difference between the stretch frequency of gas-phase \ce{H2} calculated with our chosen DFT functional and the experimental value. The translational potentials $V(d_{\text{mol--surf}}, d_{\text{tip--surf}})$ and the position-dependent wavenumbers $\nu_{\text{harm}}(d_{\text{mol--surf}}, d_{\text{tip--surf}})$ corresponding to $\omega_{\ce{H2}}$ are plotted in Fig.~\ref{fig:model-pes}.

We computed the numerically exact solutions to the 2D Schr\"{o}dinger equation corresponding to Eq.~\eqref{eq:model-pes} using the sinc-function discrete variable representation (DVR) \cite{schwartzHighaccuracyApproximation1985,Colbert1992}. The equation was solved for high coverage (0.69~ML) with \mbox{$d_{\text{tip--surf}} = 7.4\,\text{\AA}$} and \mbox{$9.4\,\text{\AA}$}. Given the large separation in the timescales of translational and vibrational motion, we also considered the vibrationally adiabatic approximation, whereby one fixes the translational coordinate $d_{\text{mol--surf}}$ in Eq.~\eqref{eq:model-pes} and solves the corresponding one-dimensional (1D) Schr\"{o}dinger equation in $r_{\text{stretch}}$, with known energy eigenvalues
\begin{equation}
    E_{\text{vib}, \upsilon}(d_{\text{mol--surf}}, d_{\text{tip--surf}}) = \hbar \omega(d_{\text{mol--surf}}, d_{\text{tip--surf}}) \qty(\upsilon + \frac{1}{2}),
\end{equation}
where $\hbar$ is the reduced Planck constant and $\upsilon=0,\,1,\,2,\,\ldots$ is the vibrational quantum number.
These are then used to construct the vibrationally adiabatic translational potential,
\begin{equation}
    \label{eq:bob}
    U_{\text{1D}, \upsilon}(d_{\text{mol--surf}}; d_{\text{tip--surf}}) = 
    V(d_{\text{mol--surf}}, d_{\text{tip--surf}}) + E_{\text{vib}, \upsilon}(d_{\text{mol--surf}}, d_{\text{tip--surf}}),
\end{equation}
for which we solve the 1D Schr\"{o}dinger equation numerically using DVR. The energy difference between the ground state solutions for $\upsilon = 0$ and $1$ is an excellent approximation to the vibrational excitation energy of the 2D model (see Table~\ref{tab:vib-adiabat}). Given the accuracy of the vibrationally adiabatic approximation, we adopt it in all subsequent calculations.

In Tables~\ref{tab:h2-quantum} and~\ref{tab:d2-quantum}, we give the vibrational transition energies computed for the H$_2$ and D$_2$ molecules at high and low coverages. For illustation purposes (see Fig.~4 in the main text), we have also computed the zero-point energies (ZPEs) and mean molecular displacements for the bare translational potentials $V(d_{\text{mol--surf}}, d_{\text{tip--surf}})$. From the computed values of $\nu$, it is apparent that if both H$_2$ and D$_2$ present a similar coverage, our model cannot explain the anomalous isotope effect observed experimentally. However, if \ce{H2} exhibits a low coverage and \ce{D2} exhibits a high coverage, the experimental trends are well accounted for [see also Fig.~4(e) in the main text]. However, it is challenging to determine the local molecular concentration near the tip apex at any measurement, and a deeper study of the vibrational frequency variation could provide indirect hints to the concentrations observed in experiment. Although we neglected the anharmonic character of the \ce{H-H} stretching coordinate, this effect should not change the qualitative discussion and will only introduce uniform redshifts in the values reported here.

\begin{table}[hbt]
    \centering
    \caption{Vibrational excitation energies calculated from the numerical solutions of the Schr\"{o}dinger equations corresponding to the 2D model in Eq.~\eqref{eq:model-pes} and the vibrationally adiabatic 1D model in Eq.~\eqref{eq:bob}. The top row indicates the values of \mbox{$d_{\text{tip--surf}}$}. The number of significant figures in the 2D solutions was estimated by varying the grid step and bounds in the DVR. The 1D solutions are converged to at least the number of significant figures shown below. %
    \label{tab:vib-adiabat}}
    \begin{tabular}{R{1.5em}*{4}{d{-1}}}
        \toprule
          &
        \multicolumn{2}{c}{$7.4~(\text{\AA})$} &
        \multicolumn{2}{c}{$9.4~(\text{\AA})$} \\ 
         \cmidrule(r){2-3} \cmidrule(l){4-5}
          & 
        \multicolumn{1}{c}{$\nu_{\text{H}_2}~(\text{cm}^{-1})$} &
        \multicolumn{1}{c}{$\nu_{\text{D}_2}~(\text{cm}^{-1})$} &
        \multicolumn{1}{c}{$\nu_{\text{H}_2}~(\text{cm}^{-1})$} &
        \multicolumn{1}{c}{$\nu_{\text{D}_2}~(\text{cm}^{-1})$} \\
        \midrule
        \text{2D} & 4073.69 & 2880.45 & 4089.44 & 2891.58 \\
        \text{1D} & 4073.685 & 2880.452 & 4089.442 & 2891.582 \\
        \bottomrule
    \end{tabular}
\end{table}

\begin{table}[htb]
    \centering
    \caption{ZPEs, mean molecule--surface displacements ($\expval{d_{\text{tip--surf}}}$) and vibrational transition frequencies~($\nu$) computed for \ce{H2} at high (0.69~ML) and low (0.06~ML) coverages, corresponding to the top and bottom rows of Fig.~\ref{fig:model-pes}, respectively. The ZPE and $\expval{d_{\text{tip--surf}}}$ are computed by solving the 1D Schr\"{o}dinger equation for \emph{only} the translational potential, $V(d_{\text{mol--surf}}, d_{\text{tip--surf}})$.
    The ZPE shown here is reported for the lowest-energy state localized in the well marked in Figs.~4(b) for the high coverage and 4(d) for the low coverage. The vibrational transition frequencies are computed for the 2D model within the vibrationally adiabatic approximation.
    The low-coverage data are used for plotting the bullets and the vertical lines on them in Fig.~4(d) and ``H$_2$ calc'' in Fig.~4(e) of the main text. 
    The low-coverage values are used for comparison with the experimental plots in Fig.~4(e).%
    \label{tab:h2-quantum}}
    \begin{tabular}{*{7}{d{-1}}}
        \toprule
        \multicolumn{1}{c}{\multirow{2}{*}{$d_{\text{tip--surf}}~(\text{\AA}) \mkern8mu$}}  &
        \multicolumn{3}{c}{$0.69~\text{ML}$} &
        \multicolumn{3}{c}{$0.06~\text{ML}$} \\ 
         \cmidrule(r){2-4} \cmidrule(l){5-7}
          & 
        \multicolumn{1}{c}{$\mkern4mu \text{ZPE}\,(\text{meV}) \mkern4mu $} &
        \multicolumn{1}{c}{$\mkern4mu \expval{d_{\text{mol--surf}}}\,(\text{\AA}) \mkern4mu $} &
        \multicolumn{1}{c}{$\mkern4mu \nu\,(\text{cm}^{-1})\mkern4mu $} &
        \multicolumn{1}{c}{$\mkern4mu \text{ZPE}\,(\text{meV}) \mkern4mu $} &
        \multicolumn{1}{c}{$\mkern4mu \expval{d_{\text{mol--surf}}}\,(\text{\AA}) \mkern4mu $} &
        \multicolumn{1}{c}{$\mkern4mu \nu\,(\text{cm}^{-1})\mkern4mu $} \\
        \midrule
        \text{11.4} & 9.7 & 3.22 & 4089.57 & 8.0 & 3.25 & 4097.62  \\
        \text{9.4}  & 9.8 & 3.22 & 4089.44 & 7.5 & 3.37 & 4101.79  \\
        \text{7.9}  & 8.5 & 3.28 & 4083.09 & 7.9 & 4.66 & 4071.55  \\
        \text{7.4}  & 7.0 & 3.41 & 4073.68 & 6.8 & 4.06 & 4063.34  \\
        \bottomrule
    \end{tabular}
\end{table}

\begin{table}[htb]
    \centering
    \caption{Same as Table~\ref{tab:h2-quantum} but for \ce{D2}. In this case, the high-coverage (0.69~ML) data are used for plotting the bullets and the vertical lines on them in Fig.~4(b) and ``D$_2$ calc'' in Fig.~4(e) of the main text.%
    \label{tab:d2-quantum}}
    \begin{tabular}{*{7}{d{-1}}}
        \toprule
        \multicolumn{1}{c}{\multirow{2}{*}{$d_{\text{tip--surf}}~(\text{\AA}) \mkern8mu$}}  &
        \multicolumn{3}{c}{$0.69~\text{ML}$} &
        \multicolumn{3}{c}{$0.06~\text{ML}$} \\ 
         \cmidrule(r){2-4} \cmidrule(l){5-7}
          & 
        \multicolumn{1}{c}{$\mkern4mu \text{ZPE}\,(\text{meV}) \mkern4mu $} &
        \multicolumn{1}{c}{$\mkern4mu \expval{d_{\text{mol--surf}}}\,(\text{\AA}) \mkern4mu $} &
        \multicolumn{1}{c}{$\mkern4mu \nu\,(\text{cm}^{-1})\mkern4mu $} &
        \multicolumn{1}{c}{$\mkern4mu \text{ZPE}\,(\text{meV}) \mkern4mu $} &
        \multicolumn{1}{c}{$\mkern4mu \expval{d_{\text{mol--surf}}}\,(\text{\AA}) \mkern4mu $} &
        \multicolumn{1}{c}{$\mkern4mu \nu\,(\text{cm}^{-1})\mkern4mu $} \\
        \midrule
        \text{11.4} & 7.1 & 3.18 & 2891.66 & 5.9 & 3.19 & 2896.83  \\
        \text{9.4}  & 7.2 & 3.18 & 2891.58 & 5.7 & 3.30 & 2899.92  \\
        \text{7.9}  & 6.3 & 3.22 & 2886.81 & 6.0 & 4.74 & 2878.13  \\
        \text{7.4}  & 5.3 & 3.33 & 2880.45 & 5.2 & 4.15 & 2872.51  \\
        \bottomrule
    \end{tabular}
\end{table}

\subsection{D. TERS Spectrum Simulation}

\begin{figure}[ht]
    \centering
    \includegraphics[width=0.7\textwidth]{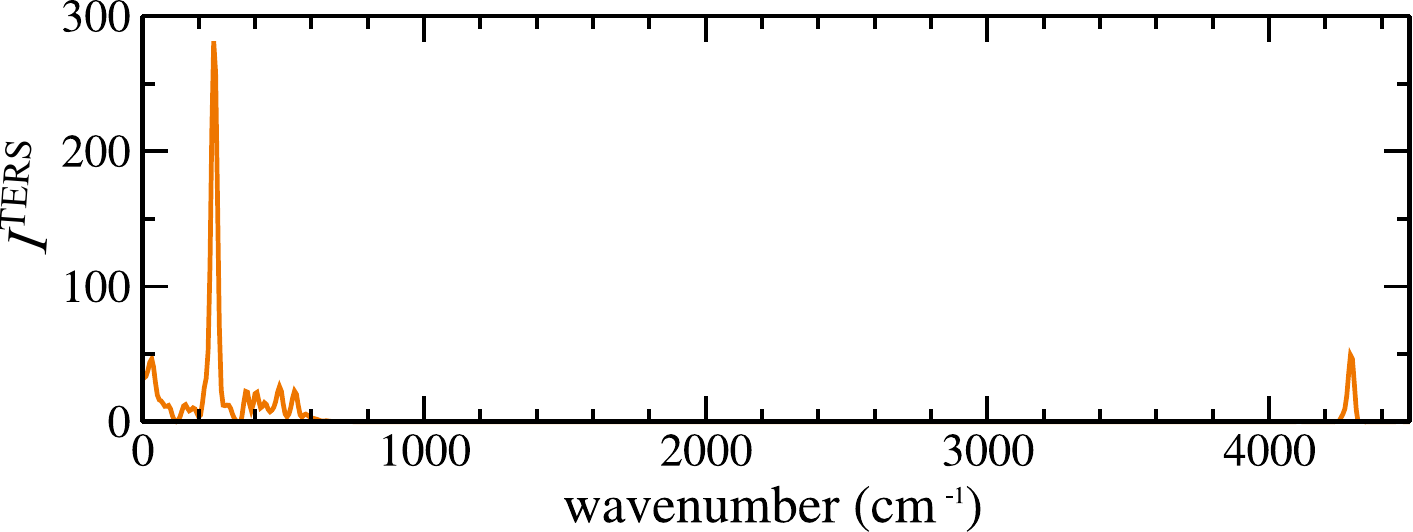}
    \caption{Simulated anharmonic TERS spectrum of H$_2$ in a tip--surface junction.
    }
    \label{fig:ters-sim}
\end{figure}

We conducted TERS simulations according to the methodology proposed in Refs.~\citep{litman2023jpcl}.
We used Tip A discussed in Ref.~\citep{cirera2022charge} and calculated the anharmonic Raman spectrum based on an \textit{ab initio} molecular dynamics (AIMD) trajectory (5 ps) of a molecular layer adsorbed on the surface and a tip apex located 4 {\AA} above the molecular layer at $T$ = 20 K.
We calculated the TERS intensity as 
$I^{\mathrm{TERS}} \propto \int \mathrm{d}t e^{i \omega t} \langle \alpha_{zz}^{\text{local}}(0) \alpha_{zz}^{\text{local}}(t)\rangle$, 
where $\alpha_{zz}^{\text{local}}$ was obtained by a local polarizability calculation at each geometry in the time-series given by the molecular dynamics.
For this part of the calculation, we only considered the free-standing molecular layer (no surface) at the positions adopted in the AIMD trajectory including the surface.
We note that the method of Ref.~\citep{litman2023jpcl} is unable to capture the plasmonic enhancement in its totality.
We do not observe any noteworthy features in the calculated spectrum and it overall agrees with the experimental spectrum.
The redshift of the rotational band in the simulation is explained by the absence of nuclear spin in the simulations, making calculations reproduce only classical rotations. These are incapable of capturing the quantization of the rotational transitions and thus do not capture the $J = 0 \to 2$ transition of \textit{para}-H$_2$, where $J$ is the rotational quantum number. This explains the redshift observed in the simulated frequency of this transition with respect to the experimental value.
This simulation is of exploratory character, as achieving statistical convergence through the use of parametrized machine-learned potentials along with accurate plasmonic enhancements is beyond the scope of this paper.

\subsection{E. FEM Simulations} \label{sec_FEM}

\begin{figure}[bth]
    \centering
    \includegraphics[height=0.27\textwidth]{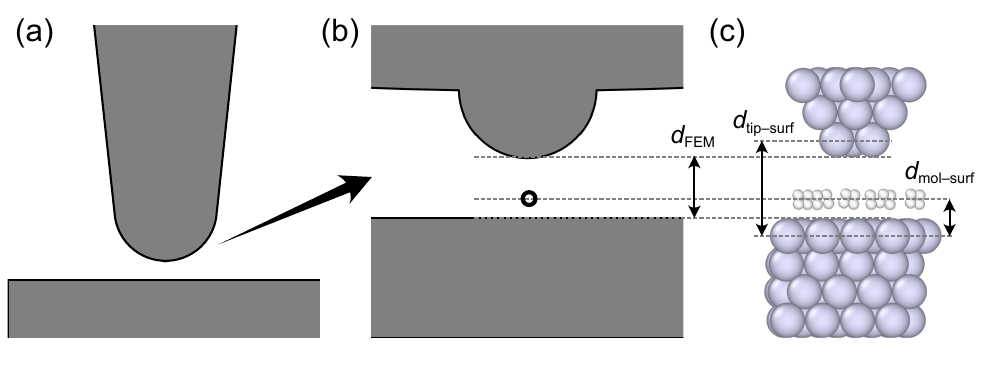}
    \caption{Side-view schemes for FEM.
    (a) Wide-area view for the FEM simulation. Gray objects correspond to Ag.
    Note that the 0.5-nm radius protrusion at the tip apex is too small to be visible.
    (b) Magnified view of the junction.
    Black open bullet indicates the sampling position of the electric field [see the black curve of Fig.~2(c) in the main text].
    (c) Side-view of the DFT model [Fig.4(a) in the main text] in the same scale as the FEM model in (b) for the comparison.
    }
    \label{fig:FEM}
\end{figure}

To evaluate the electric field in the STM junction, we performed finite element method (FEM) simulations using COMSOL Multiphysics (version 6.2 with the Wave Optics Module).
We used the three-dimensional model with an Ag tip and Ag sample plate in a vacuum  \cite{liu2019resolving}.
The tip shaft is elongated in $z$ direction with a half opening angle of 6$^\circ$ from a 30-nm radius sphere [Fig.~\ref{fig:FEM}(a)].
The tip is axisymmetrical to $z$ [see the inset of Fig.~2(c) for the $xyz$ coordinate] with a length of 300 nm.
To reproduce a plasmonic picocavity in the STM junction, at the apex of the Ag tip, an Ag half sphere with a radius of 0.5 nm is attached as an atomic-scale protrusion [Fig.~\ref{fig:FEM}(b)].
For comparison [the orange dotted curve in Fig.~2(c) of the main text], we also constructed another Ag tip model with an effective radius of 30 nm without the atomic-scale protrusion.
A $p$-polarized plane wave with $\lambda_\mathrm{exc}$ = 532 nm and an electric field $|E_0|$ of 1 V/m is incident with an angle of 55$^\circ$ relative to the $z$ axis.
The perfectly matched layers surrounding the volume are set to absorb all outgoing waves.
The dielectric constant of the Ag objects is referenced from the literature \cite{jiang2016realistic}.
The $z$ component of the electric field $E_z$ is sampled.

An Ag flat plate with a thickness of 100 nm is placed beneath the tip apex with a gap distance $d_\mathrm{FEM}$, as shown in [Fig.~\ref{fig:FEM}(b).
As shown in Table~\ref{tab:d2-quantum}, the molecular height $d_\mathrm{mol-surf}$ is calculated to be 0.32--0.33 nm for D$_2$/Ag(111) [Fig.~\ref{fig:FEM}(c)].
In the $d_\mathrm{FEM}$ definition, taking into account the atomic radius of Ag, the position of the molecule corresponds to $\sim$0.2 nm above the Ag surface plane in the FEM model [the open bullet in Fig.~\ref{fig:FEM}(b)].
Although the calculated $d_\mathrm{mol-surf}$ depends on the tip height (Table~\ref{tab:d2-quantum}), the FEM model used the constant height for the sampling point at any $d_\mathrm{FEM}$.
We confirmed that differences in the molecular position on the order of 0.1 nm do not result in qualitative differences in the $|E_z|$ curve shape.
To directly compare with the experimental TERS intensities, the electric field enhancement factor $(|E_z|/|E_0|)^4$ (Ref.~\citep{pettinger2009tip}) is displayed in Fig.~2(c).
To the field enhancement of the nearest neighboring molecule, assuming the molecule is placed in the next adsorption site of Ag(111), $x$ = 0.29 nm is used [red triangle in the inset of Fig.~2(c)].
Thus, when the point at the surface just under the tip is defined as ($x$, $y$, $z$) = (0, 0, 0),
the black and red curves shown in Fig.~2c of the main text represent the plots sampling at ($x$, $y$, $z$) = (0, 0, 0.20) nm and (0.29, 0, 0.20) nm, respectively.

As shown in Fig.~2, the gap distances with the FEM simulations $d_\mathrm{FEM} = 550$ pm to 850 pm well reproduced the Raman enhancement at the experimental tip height $z_\mathrm{exp} = -300$ to 0 pm.
Considering the Ag-atom radius, $d_\mathrm{tip-surf}$ for the simulations (Fig.~4 in the main text) is comparable to $d_\mathrm{FEM} + 0.3$~nm [Fig.~\ref{fig:FEM}(b) and (c)].
Therefore, $z_\mathrm{exp} = -300$ pm corresponds to $d_\mathrm{tip-surf} \approx 0.8$ nm, which was used for the alignment of the experimental and theoretical plots of the vibrational frequency shifts $\Delta \nu$ [Fig.~4(e)].

\section{II. Supplemental Results and Discussion}
\subsection{A. Nuclear Spin Isomers}

H$_2$ and D$_2$ have two nuclear spin isomers, \textit{ortho} and \textit{para}.
\textit{para}-H$_2$ (nuclear spin quantum number $I$ = 0) is more stable than \textit{ortho}-H$_2$ ($I$ = 1) by 14 meV \cite{wolniewicz1983x}.
\textit{para}-H$_2$ has only even $J$, whereas and \textit{ortho}-H$_2$ has only odd $J$.
Although \textit{ortho}-H$_2$ is the predominant species in the gas phase at room temperature (a temperature equilibrium composition of 75\%), the molecular adsorption on metals enhances the \textit{ortho}-to-\textit{para} conversion, leading to the dominance of the \textit{para} species on the surfaces \cite{fukutani2013physisorption}.
In the TERS spectra, we can detect only $para$-H$_2$ (Table~\ref{TabRaman});
as a reference, in the gas phase, \textit{ortho}-H$_2$ have a Raman-active rotational mode [$(\upsilon, J) = (0, 1) \to (0, 3)$] at $\sim$587 cm$^{-1}$ \cite{wolniewicz1983x,veirs1987raman}.
This implies that we detected only the physisorbed species after \textit{ortho}-\textit{para} conversion on the surface.
In the case of D$_2$, \textit{ortho} ($I$ = 0, 2) is more stable than \textit{para} ($I$ = 1) by 7.4 meV and more dominant (67\%) in the gas phase at room temperature.
Only $ortho$-D$_2$ was detected in our TERS spectra, indicating again that \textit{para}-D$_2$ was converted on the surface.

Although we introduced normal hydrogen gases (\textit{n}-H$_2$ and \textit{n}-D$_2$) into the chamber, we observed rotational/vibrational transitions only for \textit{para}-H$_2$ and \textit{ortho}-D$_2$ in the TERS spectra (Fig.~1 in the main text). 
The dominance of \textit{para}-H$_2$ on Ag(111) in the TERS spectra ($T$ = 10 K) is consistent with a previous study with high-resolution electron energy loss spectroscopy (HREELS) for H$_2$ on an Ag film (exposures of 0.5--10 L, $T$ = 10 K) \cite{avouris1982observation}, whereas some previous reports detect \textit{ortho}-H$_2$ on Ag(111).
A HREELS study on H$_2$/Ag(111) at 6 K \cite{sakurai1988ortho} shows that rotational mode peaks of both \textit{para}-H$_2$ and \textit{ortho}-H$_2$ were detected under a higher H$_2$ pressure ($9.0 \times 10^{-8}$ Torr) while \textit{para}-H$_2$ was dominantly detected under a lower pressure ($2.0 \times 10^{-9}$ Torr).
Photostimulated desorption studies \cite{fukutani2003photostimulated,niki2008mechanism} shows that at an exposure of 30 L and a temperature of $\sim$6 K, \textit{ortho}-H$_2$ and \textit{para}-D$_2$ are alive on Ag(111) on the order of hundred seconds and that the conversion of D$_2$ is slower than H$_2$.

To compare with the above-mentioned studies, we performed TERS measurements while exposing D$_2$ to a clean Ag(111) surfaces, as shown in Fig.~\ref{figS2}.
The gas pressure was set to $9.7 \times 10^{-8}$ Torr, comparable with the EELS study \cite{sakurai1988ortho}.
The \textit{ortho}-D$_2$-derived peaks appeared at $\sim$16 L and then saturated at $\sim$40 L, while \textit{para}-D$_2$ was not detected.
The weak intensity at low exposures may be due to molecular diffusion along the surface at low coverages and the weak contribution of neighboring molecules to the TERS signal [see Fig.~2(c) in the main text].
It should be noted that even if the gas pressure and exposure values are comparable, the molecular coverage varies depending on the chamber configuration.
Nevertheless, we confirmed that \textit{ortho}-H$_2$ and \textit{para}-D$_2$ were not detected at any pressures and exposures we used (0.8--2 $\times 10^{-7}$ Torr; 10$^2$--10$^3$ L).
The HREELS observation of the metastable isomer at 6 K only at high exposures \cite{sakurai1988ortho} may be attributed to be hydrogen multilayers, where the second layer does not interact with the Ag surface. 
For our LT-TERS measurement, in contrast, only a monolayer would be formed at the higher temperature, 10 K. 
Another possibility that the metastable isomer were not detected in TERS is that the localized surface plasmonic resonance (LSPR) in the gap is accelerating the \textit{para}-to-\textit{ortho} conversion of the molecule under the tip.
Although far-field 532-nm irradiation did not affect the conversion rate \cite{niki2008mechanism}, the strong electric field localized in the STM junction could interact with the nuclear spin of the adsorbate.


\begin{table}[h]
\centering
\caption{Frequencies of rotation/vibrotation modes of molecular hydrogen detected experimentally (in cm$^{-1}$).
From the TERS results in this study, data from a wide-frequency-range spectra [``low res.''; Fig.~1(a) in the main text] and from spectra with higher energy resolution [``high res.''; Figs.~3(a)--3(d)] are displayed.
Since the vibration frequency of H$_2$/Ag(111) is largely redshifted by the tip proximity, the values at a far tip distance [a relative tip height of $-$110 pm in Figs.~3(a)--3(d)] are shown as the high resolution data. }
\label{TabRaman}
\begin{tabular}{lrrrrrr}
\hline
\multicolumn{1}{c}{System} & \multicolumn{2}{c}{On Ag(111)}                                               & \multicolumn{1}{c}{On Ag film} & \multicolumn{1}{c}{On Cu(001)} & \multicolumn{2}{c}{Gas phase}                           \\
\multicolumn{1}{c}{Method} & \multicolumn{1}{c}{TERS (low res.)} & \multicolumn{1}{c}{TERS (high res.)} & \multicolumn{1}{c}{HREELS$^a$}    & \multicolumn{1}{c}{HREELS$^b$}   & \multicolumn{1}{c}{Raman$^c$} & \multicolumn{1}{c}{Raman$^d$} \\
\hline
H$_2$                         &                                      &                                       &                                &                               &                            &                            \\
rot                        & 351                                  & 348                                   & 395                            & 363                           & 354.5                      & 354.365                    \\
vib                        & 4121                                 & 4134                                  & 4178                           & 4178                          & -                          & 4161.200                   \\
rot+vib                    & 4443                                 & -                                     & 4533                           & 4517                          & -                          & 4497.848                   \\ 
\hline
D$_2$                         &                                      &                                       &                                &                               &                            &                            \\
rot                        & 169                                  & 172                                   & -                              & 177                           & 178.8                      & 179.108                    \\
vib                        & 2967                                 & 2968                                  & -                              & 3000                          & 2996.1                     & 2993.6~~~                     \\
rot+vib                    & 3133                                 & -                                     & -                              & 3170                          & 3166.3                     & -                          \\
\hline
\end{tabular}
\begin{minipage}{0.7\columnwidth}
\begin{flushleft}
{$^a$ Ref.~\citep{avouris1982observation}.}\\
{$^b$ Ref.~\citep{andersson1982observation}.}\\
{$^c$ Ref.~\citep{teal1935raman}.}\\
{$^d$ Ref.~\citep{veirs1987raman}.}

\end{flushleft}
\end{minipage}
\end{table}

\subsection{B. Excitation Wavelength Dependence}
\label{sec:wavelengthdep}

\begin{figure*}[bth]
\includegraphics[width=11cm]{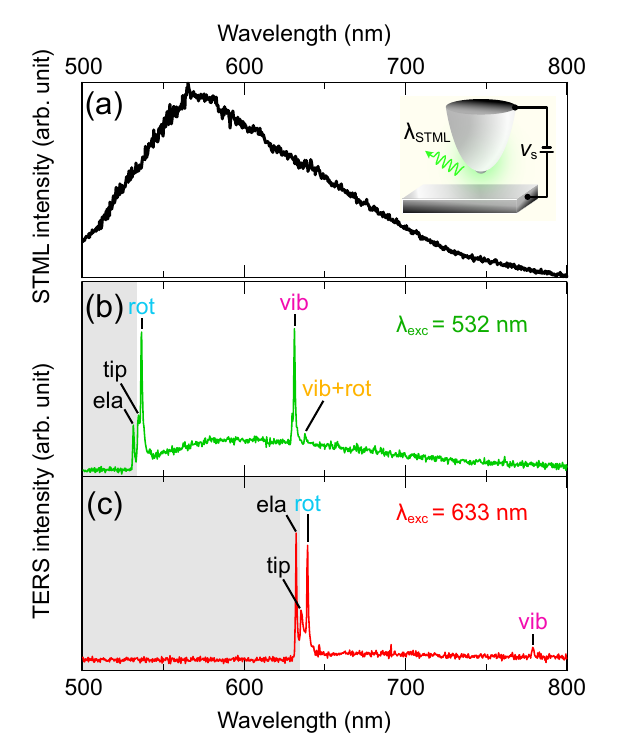}
\caption{\label{figS3}
Correlation between Raman scattering and plasmonic resonance in the STM junction.
(a) STML recorded over a clean Ag(111) surface with an Ag tip ($V_\mathrm{s} = 3$ V, $I_\mathrm{t} = 5$ nA, $T$ = 10 K).
The inset shows the schematic of the STML experiment.
The horizontal axis is the luminescence wavelength $\lambda_\mathrm{STML}$.
After recording this spectrum, the surface was exposed to the D$_2$ gas.
(b and c) TERS spectra of D$_2$/Ag(111) with $\lambda_\mathrm{exc}$ = 532 and 633 nm, respectively ($V_\mathrm{s} = 10$ mV, $I_\mathrm{t} = 1$ nA), using the Ag tip as in (a).
The horizontal axis is indicated by wavelength $\lambda_\mathrm{Raman}$ instead of the Raman shift.
The gray background represents the long-pass filtered region.
``ela'' represents the peak of the elastic (i.e., Rayleigh) scattering with its intensity suppressed by the filter.
}
\end{figure*}

To verify the contribution of LSPR to the TERS intensity, we examined the excitation wavelength ($\lambda_\mathrm{exc}$) dependence.
Figure~\ref{figS3}(a) shows a STML spectrum recorded over a clean Ag(111) surface with an Ag tip.
The LSPR results in a broad spectral feature in the visible range with a maximum at $\sim$560 nm.
After recording the STML, we exposed the surface to the D$_2$ gas, conducting TERS measurements with $\lambda_\mathrm{exc}$ = 532 and 633 nm [Figs.~\ref{figS3}(b) and \ref{figS3}(c), respectively].
The peak intensity of each mode significantly changes with $\lambda_\mathrm{exc}$, whereas the Raman shift value is unchanged.
The vibrational intensity is weakened with $\lambda_\mathrm{exc}$ = 633 nm, indicating the corresponding Raman scattering is not efficiently enhanced by the LSPR because the peak position (a wavelength of the Raman scattering $\lambda_\mathrm{Raman}$ of 779 nm) is located in the weak resonance range of the LSPR [Fig.~\ref{figS3}(a)].
In contrast, both the rotational and vibrational peaks become intense with $\lambda_\mathrm{exc}$ = 532 nm because they are more resonant with the LSPR.

\subsection{C. Tip-Height Dependence of Peak Intensity and Width}

\begin{figure*}[bth]
\includegraphics[width=11cm]{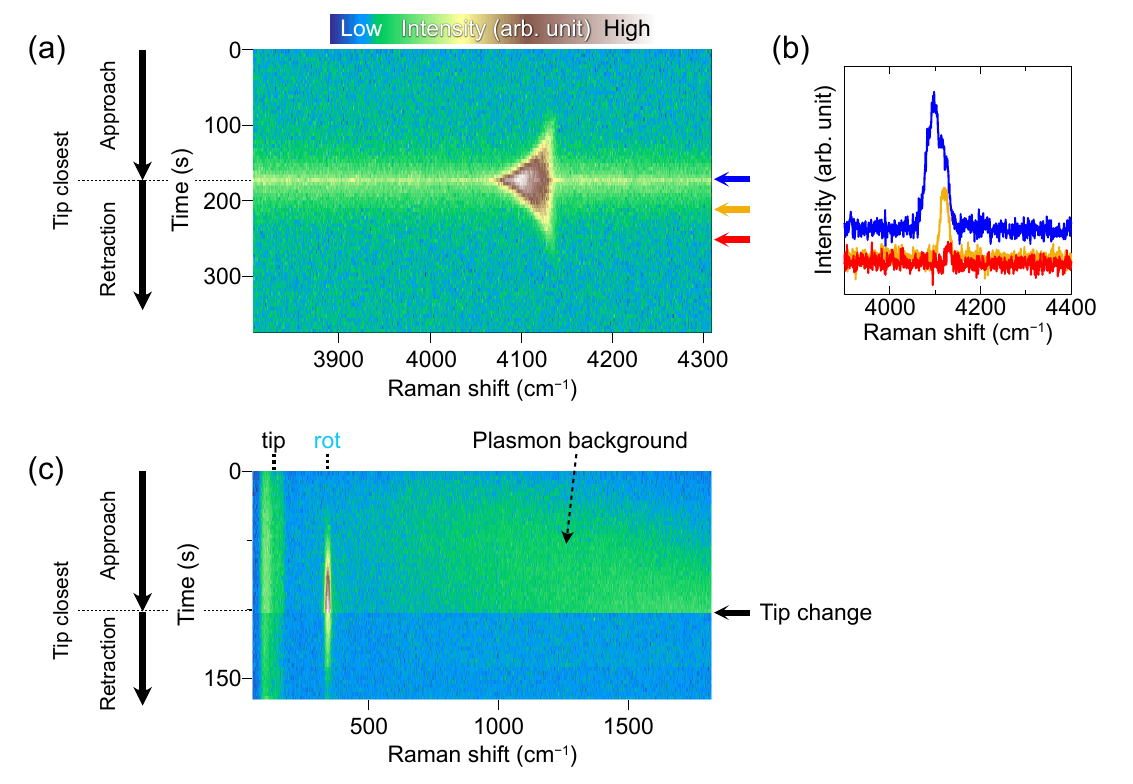}
\caption{\label{figS11}
(a) Full-scale waterfall plot of the high-resolution TERS spectrum for the H$_2$ ``vib'' mode shown in Fig.~3(b) of the main text.
(b) TERS spectra of H$_2$/Ag(111), corresponding to the horizontal line profiles in (a) indicated by the arrows in the same color.
No vertical offset is applied to the graph; the blue spectrum has a higher constant background as shown in (a).
(c) Waterfall plot of the high-resolution TERS spectrum for the H$_2$ ``rot'' mode, where a sudden drop of the TERS intensity was observed.
}
\end{figure*}

To obtain the waterfall plots shown in Fig.~3 of the main text, we conducted tip approach-and-retraction procedures while recording the TERS spectra \cite{cirera2022charge,liu2023inelastic}.
Figure~\ref{figS11}(a) is the full dataset for the H$_2$ ``vib'' mode shown in Fig.~3(b) of the main text.
During recording TERS spectra, the tip height was lowered stepwise and then the tip was retracted from the surface.
The symmetrical waterfall plot with respect to the closest tip height indicates the non-destructive measurement in the tip approach-and-retraction process.
Figure~\ref{figS11}(b) shows the horizontal line profiles of the waterfall plot for the H$_2$ ``vib'' mode at the points indicated by the arrows in the same color.
The three spectra indicates the redshift and broadening of the H$_2$ ``vib'' peak as the gap distance decreases.

During recording the waterfall TERS plot, we occasionally observed a sudden intensity change caused presumably by the modification of the tip apex structure due to the strong tip--surface interaction at the small gap.
Figure~\ref{figS11}(c) shows an example, where a sudden drop of the TERS intensity was observed.
The modification of the plasmon background intensity implies that the atomic-scale structure was changed by the strong attractive force exerted when the tip was brought close to the surface.
At the same time as the plasmon feature, the D$_2$ ``rot'' peak was also dropped, while the frequency of the ``rot'' mode is unchanged.
This observation is in good agreement with the enhancement mechanism described in the main text (Fig.~2); the picocavity-derived electric field plays a crucial role in the TERS intensity of physisorbed hydrogen molecules.

\begin{figure*}[tbh]
\includegraphics[width=10cm]{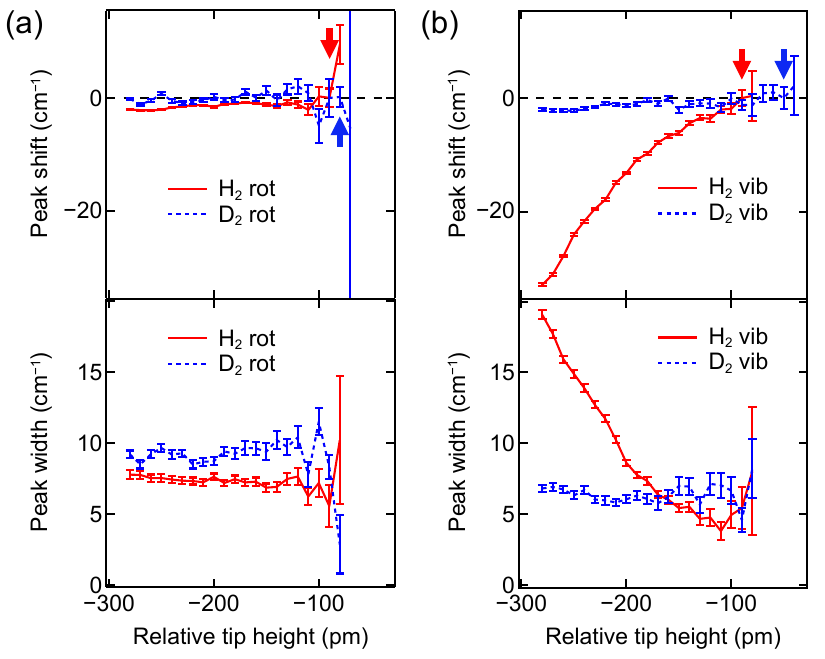}
\caption{\label{figS12}
(a) Tip-height dependence of the peak position (top panel) and  the full width at half maximum (bottom panel) for H$_2$ ``rot'' (red curve) and D$_2$ ``rot'' (blue curve).
The error bars indicate the standard deviation of the fitting.
(b) Same as in (a) but for H$_2$ ``vib'' (red curve) and D$_2$ ``vib'' (blue curve) by the fitting of the spectra shown in Fig.~3(c) and (d) in the main text, respectively.
The arrows indicate the reference points with the errors below 2 cm$^{-1}$.
}
\end{figure*}

To determine the experimental peak position shift $\Delta \nu$ as a function of the tip height $z_\mathrm{exp}$ in Fig.~4(e) of the main text, the line profile of the waterfall plots at each tip height [Fig.~3 in the main text; see also Fig.~\ref{figS11}(b)] was fitted by a Gaussian peak with a constant background.
The top and bottom panels of Fig.~\ref{figS12} shows the peak position and width, respectively, as a function of the tip height for the rotational (a) and vibrational (b) modes.
To display the experimental peak-shift values $\Delta \nu (z_\mathrm{exp}) = \nu (z_\mathrm{exp}) - \nu(z_\mathrm{ref})$, the reference point $z_\mathrm{ref}$ should be specified.
Ideally $z_\mathrm{ref}$ is the furthest tip height at which the peak was detected.
However, since using a too weak peak as the reference value $\nu(z_\mathrm{ref})$ causes large errors in the $\Delta \nu (z_\mathrm{exp})$ plot, we adopted as $z_\mathrm{ref}$ the tip height at which the error value below 2 cm$^{-1}$ based on the peak fitting.
The error bars in Fig.~\ref{figS12} indicate the standard deviation of the fitting.
As the tip approaches, the peak intensity increases, reducing the error.
As shown in the arrows in Fig.~\ref{figS12}, we set $z_\mathrm{ref} = -90$, $-80$, $-90$, $-50$ pm for H$_2$ rot, D$_2$ rot, H$_2$ vib, and D$_2$ vib, respectively, determining the origin of each peak shift.
For the H$_2$ and D$_2$ ``vib'' modes [top panel of Fig.~\ref{figS12}(b)], the $\Delta \nu$ plots at $z_\mathrm{exp} \leq z_\mathrm{ref}$ were used as the experimental curves shown in Fig.~4(e) of the main text. 
In Fig.~4(e) of the main text, the experimental tip height $z_\mathrm{exp} = -280$ pm is aligned to the tip--sample distance $d_\mathrm{tip-surf} = 0.74$ nm for the calculations.
Although the accurate determination of the experimental tip--sample gap distance is difficult, this alignment is reasonable; 
as described in Sec.~I-E, $z_\mathrm{exp} = -300$ pm is comparable with $d_\mathrm{FEM} = 550$ pm and  $d_\mathrm{tip-surf} \approx 0.8$ nm.

As shown in Fig.~\ref{figS12}(a), the frequency and width of the H$_2$ and D$_2$ ``rot'' peaks are almost unchanged by the tip.
In contrast, the bottom panel Fig.~\ref{figS12}(b) shows the strong peak broadening for the H$_2$ ``vib'' mode [see also Fig.~\ref{figS11}(b)] whereas the peak width for D$_2$ ``vib'' is almost unchanged.
A possible origin of the peak broadening is the anharmonicity, which can modify the lifetime of molecular vibrations \cite{persson1989vibrational}.
It is also entirely possible that if neighboring molecules influence the experimental signal [see Fig.~2(c) in the main text], they would lead to the high-frequency shoulder of the vibration transition for H$_2$, as these molecules would still lie on the surface [Fig.~3(c)] and would therefore present a higher $\nu$ than the ones on the tip, which instead provide the most enhanced signal; however, this effect may not be pronounced because the coverage of H$_2$ is less than that of D$_2$ (see Sec.~I-C). 
Another possible origin if the movement of the molecule in the junction; because the double-well potential barrier is very small [see Fig.~4(d) in the main text], the H$_2$ molecule near the tip apex can move to a metastable site near the surface or adjacent adsorption sites on the surface \cite{gupta2005strongly} by a small voltage bias or plasmonic field applied in the STM junction.
This movement instantaneously occurring during the acquisition time of the Raman spectrum can also modify the peak shape of the vibrational mode.
We expect that the coexistence of these factors leads to the characteristic peak width depending on the tip height.

\end{document}